\documentclass[aps,floatfix,superscriptaddress,nofootinbib]{revtex4-2}

\usepackage{amsmath,amssymb,bm}

\usepackage{graphicx}
\usepackage{booktabs}
\usepackage{dcolumn}
\usepackage{braket}

\usepackage{subcaption}
\usepackage[colorlinks=true,citecolor=blue,linkcolor=blue,urlcolor=blue]{hyperref}

\begin{document}

\title{Quantum Machine Learning for Complex Systems}

\author{Vinit Singh}
\affiliation{Department of Electrical and Computer Engineering, North Carolina State University, Raleigh, NC 27606}

\author{Amandeep Singh Bhatia}
\affiliation{Department of Electrical and Computer Engineering, North Carolina State University, Raleigh, NC 27606}

\author{Mandeep Kaur Saggi}
\affiliation{Department of Electrical and Computer Engineering, North Carolina State University, Raleigh, NC 27606}

\author{Manas Sajjan}
\affiliation{Department of Electrical and Computer Engineering, North Carolina State University, Raleigh, NC 27606}

\author{Sabre Kais}
\email{skais@ncsu.edu}
\affiliation{Department of Electrical and Computer Engineering, North Carolina State University, Raleigh, NC 27606}

\begin{abstract}

{\color{black} 

Quantum machine learning (QML) is rapidly transitioning from theoretical promise to practical relevance across data-intensive scientific domains. In this Review, we provide a structured overview of recent advances that bridge foundational quantum learning principles with real-world applications. 
We survey foundational QML paradigms, including variational quantum algorithms, quantum kernel methods, and neural-network quantum states, with emphasis on their applicability to complex quantum systems.
We examine neural-network quantum states as expressive variational models for correlated matter, non-equilibrium dynamics, and open quantum systems, and discuss fundamental challenges associated with training and sampling.
Recent advances in quantum-enhanced sampling and diagnostics of learning dynamics, including information-theoretic tools, are reviewed as mechanisms for improving scalability and trainability. The Review further highlights application-driven QML frameworks in drug discovery, cancer biology, and agro-climate modeling, where data complexity and constraints motivate hybrid quantum-classical approaches.
We conclude with a discussion of federated quantum machine learning as a route to distributed, privacy-preserving quantum intelligence.
Overall, this Review presents a unified perspective on the opportunities and limitations of QML for complex systems.

}


\end{abstract}

\maketitle

\section{Introduction}

Complex systems arise across a wide range of scientific domains, including condensed-matter and chemical systems, biological and social networks, and climate and agro-environmental dynamics, where macroscopic behavior emerges from the collective interaction of many microscopic degrees of freedom.
Such systems are typically characterized by high dimensionality, strong correlations, nonlinear interactions, and heterogeneous structure, which together give rise to rich phenomena such as multiscale organization, criticality, and emergent phases.
Classical approaches to modeling and learning in these settings typically proceed by constructing reduced descriptions, approximate generative models, or sampling-based representations that aim to capture dominant correlations while discarding microscopic detail.
In practice, these strategies rely on assumptions such as weak correlations, locality, separability, or efficient sampling from the underlying state or data distribution.
While successful in many regimes, they can break down in strongly correlated, frustrated, or highly heterogeneous settings, where reduced descriptions become inaccurate and sampling procedures suffer from slow mixing, large variance, or sensitivity to model assumptions.
Quantum many-body systems represent an extreme instance of these challenges, as the exponential growth of Hilbert space with system size renders exact simulation infeasible and exposes fundamental limitations of classical approximation and Monte Carlo methods \cite{Troyer2005ComputationalComplexity,Binder1986MonteCarlo}.
Related difficulties also arise in data-driven domains such as healthcare and Earth system science, where data are noisy, heterogeneous, and often non-IID, complicating efforts to construct reliable models from limited observations \cite{Esteva2019HealthcareAI,Reichstein2019DeepLearningClimate,Rudin2019MLRisks}.

Quantum machine learning (QML) has emerged as an interdisciplinary research area at the intersection of quantum information science and modern learning theory, motivated by the possibility that quantum resources can offer new approaches to representation, sampling, and inference in complex learning tasks \cite{biamonte2017quantum,Dunjko2018MLReview,Ciliberto2018QMLPerspective}.
Early visions of QML were closely tied to the prospect of quantum speedups for canonical machine learning primitives, including linear algebra routines, principal component analysis, clustering, and support vector machines, often relying on amplitude encoding and algorithmic subroutines such as phase estimation and Hamiltonian simulation.
These works established important theoretical benchmarks, demonstrating that quantum computers could, under specific data access models, offer polynomial or exponential improvements in computational complexity.
At the same time, they exposed practical challenges related to data loading, noise sensitivity, and error accumulation, which complicate the direct realization of such advantages on near-term quantum hardware \cite{Schuld2018Supervised,Dunjko2018MLReview,Ciliberto2018QMLPerspective}.
As the field matured, the focus of QML shifted from purely algorithmic speedups toward the design of learning models and workflows compatible with noisy intermediate-scale quantum (NISQ) devices.
This transition gave rise to hybrid quantum-classical approaches, in which parameterized quantum circuits or quantum subroutines are embedded within classical optimization loops and evaluated under realistic resource constraints \cite{Schuld2018Supervised,cerezo2021variational}.
Within this paradigm, QML is no longer viewed as a single algorithmic framework, but rather as a collection of complementary modeling strategies that differ in how quantum resources are used for representation, optimization, and inference.
Crucially, the relevance of QML to complex systems does not hinge on strong claims of universal quantum advantage.
Instead, quantum models are increasingly valued for the inductive biases and computational primitives they introduce, such as high-dimensional Hilbert space representations, non-classical correlations, and quantum-enhanced sampling dynamics, which can be naturally aligned with the structure of correlated systems and high-dimensional data.
Today, QML encompasses a broad spectrum of approaches spanning variational quantum algorithms, quantum kernel and feature-map methods, neural-network quantum states, and learning-theoretic analyses of trainability and generalization.
These paradigms have been explored across a growing range of application domains, including quantum many-body physics, quantum chemistry and materials science, drug discovery, biomedical data analysis, and climate and agro-environmental modeling \cite{Sajjan2022Quantum,JACS2021}.
Rather than constituting a unified solution, QML provides a flexible toolkit whose effectiveness depends on the interplay between representation, sampling, optimization, and deployment constraints.
This perspective motivates the organization of the present Review, which surveys how different QML paradigms address these interconnected aspects, and examines their strengths and limitations in the context of complex systems.

\subsection{Major Paradigms in Quantum Machine Learning}
\label{subsec:major-paradigms}

\textbf{Variational circuit-based learning.}
The emergence of noisy intermediate-scale quantum (NISQ) devices catalyzed a transition from fully quantum learning algorithms toward hybrid quantum-classical frameworks.
Within this paradigm, variational quantum algorithms (VQAs) formulate learning, simulation, and optimization tasks as minimization problems over parameterized quantum circuits (PQCs), with classical optimizers updating circuit parameters based on measurement outcomes \cite{cerezo2021variational}.
This approach enabled some of the earliest practical demonstrations of quantum-assisted learning and simulation, most notably through the variational quantum eigensolver (VQE), which approximates ground-state properties of quantum systems using shallow circuits and iterative optimization \cite{Peruzzo2014VQE}.

Building on this framework, quantum neural networks (QNNs) were introduced as circuit-based analogues of classical neural networks, with tunable gates acting as trainable parameters.
QNNs have been explored for supervised learning, generative modeling, and quantum data processing, and have been implemented on both simulators and available quantum hardware.
However, extensive theoretical and numerical studies have revealed that many variational circuit architectures suffer from barren plateaus, where gradients vanish exponentially with system size or circuit depth, rendering training ineffective \cite{McClean2018BarrenPlateau}.
Subsequent work has investigated how circuit expressivity, entanglement structure, and connectivity influence gradient statistics, highlighting fundamental trade-offs between representational power and trainability \cite{Holmes2022Connectivity}.
These insights have motivated the development of structured ansätze, locality-preserving circuits, problem-informed initialization strategies, and alternative optimization techniques, although scalability remains an open challenge in highly correlated or rugged learning landscapes.

\textbf{Neural-network quantum states.}
Neural-network quantum states (NQS) constitute a representational paradigm within QML in which quantum many-body wavefunctions are parameterized directly by classical machine learning models.
The seminal demonstration that restricted Boltzmann machines (RBMs) can approximate ground states of interacting spin systems initiated a broad research program on neural-network-based wavefunction representations \cite{Carleo2017Solving,Carleo2016Solving}.
Subsequent theoretical and numerical studies extended these ideas to frustrated magnets, fermionic systems, topological phases, and models with long-range interactions, establishing NQS as flexible variational ansätze for strongly correlated quantum matter \cite{Medvidovic2024Neural,Yang2019Approximating,Kim2024Neural}.
NQS combine ideas from variational Monte Carlo, statistical mechanics, and representation learning, providing a compact way to encode high-order correlations and entanglement.
A wide range of architectures, including deep neural networks, convolutional models, and autoregressive constructions, have been proposed to enhance expressivity and sampling efficiency.
Despite their flexibility, NQS-based methods typically rely on Monte Carlo estimation of observables and gradients, making training performance sensitive to autocorrelation times, mixing efficiency, and estimator variance, particularly in systems with rugged or multi-modal configuration spaces.

\textbf{Kernel- and embedding-based quantum learning.}
Quantum kernel methods provide an alternative approach to QML that avoids explicit variational training of quantum parameters.
In this paradigm, quantum circuits define implicit feature maps that embed data into high-dimensional Hilbert spaces, while learning is performed through kernel evaluation and classical optimization \cite{havlivcek2019supervised,Schuld2019Neural}.
Quantum kernels can induce complex decision boundaries using relatively shallow circuits, making them attractive for learning tasks in low-data or structured regimes.

At the same time, the practical performance of quantum kernels depends critically on the efficiency and accuracy of kernel estimation, which can be affected by noise, shot statistics, and circuit depth.
Recent studies have investigated the expressivity, generalization properties, and classical simulability of quantum kernels, clarifying the regimes in which they may offer advantages over classical counterparts.
This body of work suggests that quantum kernel methods are most effective when the structure of the feature map aligns closely with the underlying data distribution, rather than as universal replacements for classical kernel techniques.

\textbf{QML for quantum simulation and scientific modeling.}
A major application-driven paradigm of QML focuses on modeling and controlling complex quantum systems, including quantum many-body phases, correlated molecules, and quantum materials.
Foundational reviews have articulated opportunities at the interface of quantum machine learning and quantum simulation, while subsequent work has expanded these approaches across condensed-matter and chemical contexts \cite{Sajjan2022Quantum,JACS2021,Melnikov2023Quantum,Matthew2025Quantum}.

In quantum chemistry and materials science, QML techniques have been used to approximate molecular wavefunctions, reduced density matrices, and effective energy landscapes, as well as to assist in state preparation and observable estimation for electronic-structure problems.
Hybrid quantum-classical workflows have explored the use of parameterized circuits and learned representations for periodic systems and band-structure calculations on near-term devices \cite{Sureshbabu2021Implementation,Sajjan2022Quantum}.
Across these applications, a central difficulty lies in exploring highly structured configuration spaces associated with strong electronic correlations, distorted geometries, and extended systems, where classical sampling and optimization strategies can become increasingly inefficient \cite{Sajjan2022Quantum,Matthew2025Quantum,chemicalAI2024Explainable}.

\textbf{QML for non-equilibrium and open-system dynamics.}
While early applications of QML focused predominantly on ground-state properties, a growing body of work has extended machine-learning-based ans\"atze to real-time and far-from-equilibrium dynamics.
In the neural-network quantum state (NQS) framework, time evolution is typically formulated through a time-dependent variational principle (TDVP), in which the Schr\"odinger equation is projected onto the manifold of parameterized wavefunctions.
This reduces unitary dynamics to differential equations for the network parameters, allowing interacting many-body systems to be evolved approximately in real time.
Structure-preserving and symplectic integrators have been developed to improve numerical stability and mitigate drift in conserved quantities \cite{Gutierrez2022Real}, while autoregressive and deep architectures have been explored to enhance scalability and control error growth during long-time evolution \cite{Donatella2023AutoregressiveDynamics}.
Complementary progress has emerged in circuit-based hybrid algorithms for dynamical simulation.
Variational quantum simulation (VQS) recasts time evolution, both real and imaginary, as an optimization problem over parameterized quantum circuits, enabling approximate propagation using shallow circuits compatible with noisy intermediate-scale quantum (NISQ) devices \cite{Yuan2019Theory,Endo2020GeneralProcesses}.
In this framework, expectation values of the Hamiltonian and its derivatives are estimated on quantum hardware, while classical optimizers update circuit parameters to approximate the target evolution.

Open quantum systems present additional challenges, as their evolution is governed by master equations such as the Lindblad equation, which describes non-unitary dynamics arising from coupling to an environment.
To address this setting, both NQS-based and variational circuit-based methods have been adapted to represent mixed states and steady states.
Neural-network variational Monte Carlo techniques have been used to approximate non-equilibrium steady states and dissipative many-body dynamics by representing density operators via purification or stochastic unraveling schemes \cite{Vicentini2019OpenSteadyStates,Nagy2019OpenVMC}.
In parallel, variational quantum algorithms have been proposed to simulate Lindblad evolution directly on quantum hardware and to target steady states through cost functions derived from the master equation \cite{Liu2021OpenSteadyStates,Chen2024AdaptiveOpen,Watad2023LindbladVQA}.
Together, these developments extend QML beyond equilibrium physics, establishing a methodological foundation for studying driven, dissipative, and far-from-equilibrium quantum systems within hybrid quantum-classical frameworks.

\textbf{Theoretical foundations and scalability of QML.} 
Across QML paradigms, optimization dynamics and sampling complexity impose fundamental constraints on scalability.
Slow mixing, noisy gradient estimates, and ill-conditioned natural-gradient metrics can impede convergence even when expressive models exist, affecting both circuit-based and neural-network-based approaches.
These challenges have motivated the development of improved estimators, structured ansätze, and diagnostics that probe the geometry of the learning landscape.

Recent work has sought to understand QML training through the lens of learning theory and quantum information.
Measures of expressivity, entanglement growth, and information propagation have been linked to trainability and generalization behavior.
In particular, tools originally developed to study information scrambling, such as out-of-time-order correlators, have been proposed as diagnostics of information flow and landscape structure in QML models \cite{Sajjan2023ImaginaryOTOC}.
Together, these efforts emphasize that scalable QML requires not only expressive representations, but also principled mechanisms for stable optimization in high-dimensional correlated systems.

This Review surveys avenues in quantum machine learning for complex systems. We focus on a set of closely related approaches that address practical challenges in training, understanding, and deploying QML models. The remainder of the Review is organized as follows.
In Sec.~\ref{sec:Q-VMC}, we examine quantum-enabled variational Monte Carlo frameworks for training neural-network quantum states, addressing sampling bottlenecks that limit classical optimization in strongly correlated systems.
In Sec.~\ref{sec:OTOC}, we survey emerging diagnostic and theoretical tools for understanding the learning dynamics of QML models, with particular focus on information-theoretic and scrambling-inspired measures.
Sec.~\ref{sec:QML-Bio} reviews application-driven QML frameworks in domains such as drug discovery, cancer biology, and agro-climate modeling, where high-dimensional, structured, and often limited data pose stringent challenges for learning algorithms.
Finally, in Sec.~\ref{sec:FQML}, we discuss federated quantum machine learning as a pathway toward distributed and privacy-preserving quantum intelligence, enabling collaborative learning across heterogeneous and geographically distributed data sources.

\begin{figure}[!ht]
    \centering
    \includegraphics[width=0.5\linewidth]{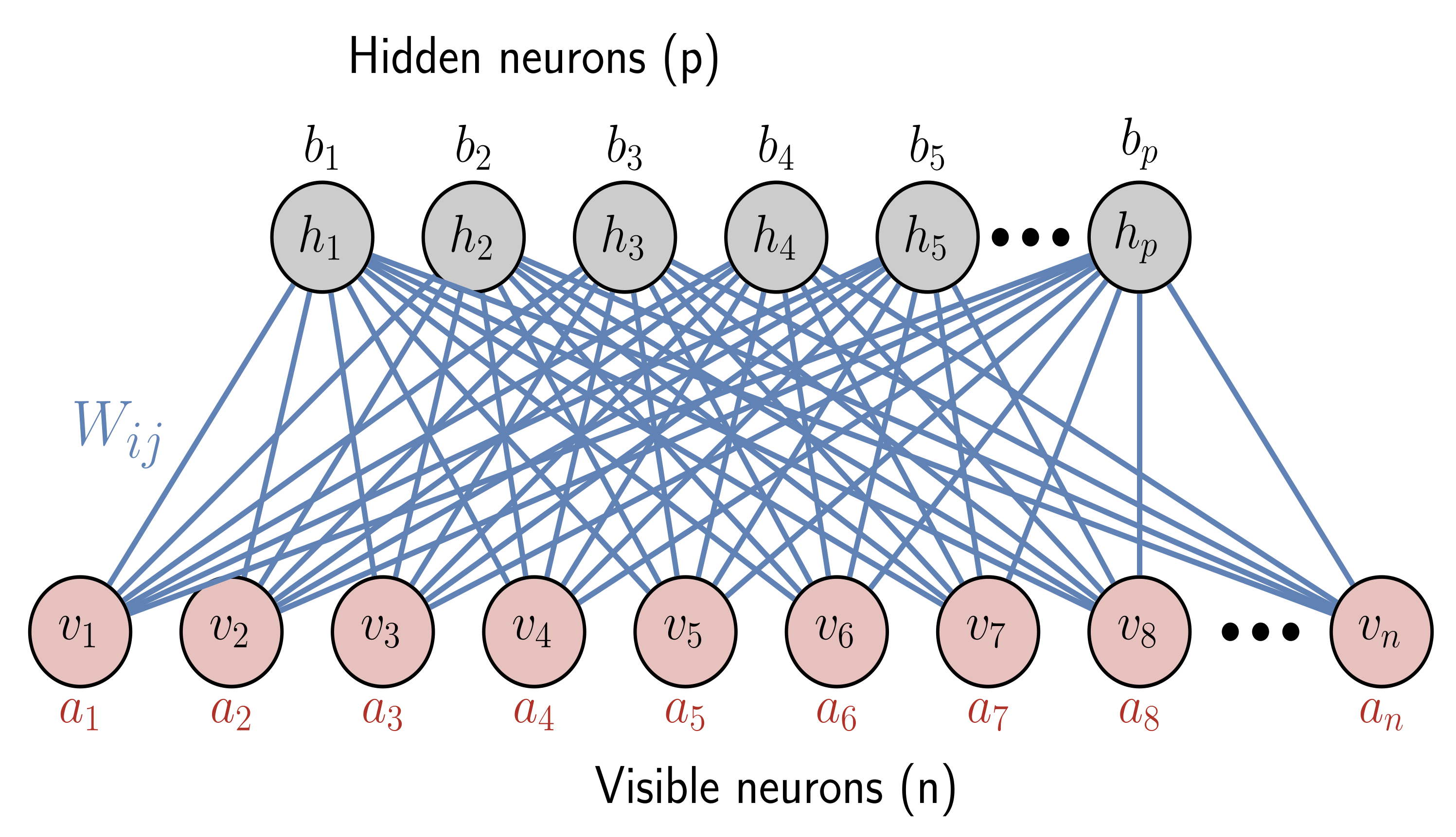}
    \caption{A schematic of a Restricted Boltzmann Machine (RBM), a widely used neural-network quantum state. 
    The learner network is represented by a bipartite graph \( G = (V, E) \), consisting of a set of hidden neurons 
    \( \{ h_j \}_{j=1}^{p} \), shown in gray and associated with bias vector \( \vec{b} \), and a set of visible neurons 
    \( \{ v_i \}_{i=1}^{n} \), shown in red and associated with bias vector \( \vec{a} \). 
    The inter-layer connections are encoded by the weight matrix \( \vec{W} \), with elements \( W_{ij} \), shown in blue. 
    The full parameter set \( \vec{X} = (\vec{a}, \vec{b}, \vec{W}) \) is optimized during the training of the network.
    }
    \label{fig:RBM}
\end{figure}

\section{Quantum-enabled variational Monte Carlo algorithm for training neural-network quantum states}
\label{sec:Q-VMC}

Although NQS have proven remarkably expressive across a wide range of correlated quantum systems, their practical performance in large-scale simulations is often limited by the cost of classical sampling. Conventional variational Monte Carlo (VMC) optimization relies on classical Markov chain Monte Carlo (MCMC) sampling, whose mixing times and autocorrelation lengths can grow exponentially with system size. This challenge becomes especially severe in frustrated or multireference regimes, where the probability distribution encoded by the NQS wave function develops multiple deep local minima separated by high-energy barriers. As a result, classical samplers struggle to explore configuration space efficiently, making the optimization landscape rugged and slowing convergence. This sampling bottleneck represents one of the central obstacles to scaling NQS-based electronic structure and condensed matter simulations beyond modest system sizes.

In our recent work \cite{Sajjan2023Polynomial}, we introduced a quantum-enabled variational Monte Carlo (Q-VMC) framework that replaces the classical sampling step with a shallow quantum circuit implementing the dynamics of an auxiliary Ising-like Hamiltonian designed to mimic the structure of the desired probability distribution. The key conceptual step is the construction of a surrogate Hamiltonian $h_1$ whose Gibbs distribution approximates the  probability distribution of the NQS $|\psi_{\theta}(x)|^{2}$. 

\begin{figure}[h!]
    \centering
    \includegraphics[width=0.66\linewidth]{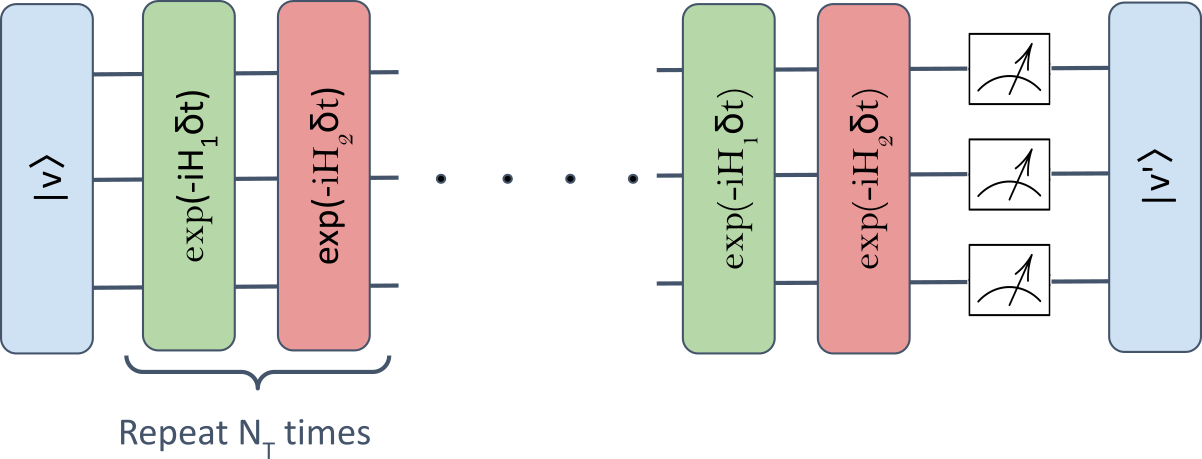}
    \caption{Trotterized quantum circuit implementing the quantum-enhanced proposal distribution. The proposal transition probability is defined as
    \(
    P_{\mathrm{prop}}\!\left(\vec v^{(i+1)} \mid \vec v^{(i)}\right)
    = \left|\langle \vec v^{(i+1)} | U(\tau,\gamma) | \vec v^{(i)} \rangle\right|^2 ,
    \)
    where the unitary evolution operator is
    \(
    U(\tau,\gamma) = e^{-i\tau\left[\gamma, h_1 + (1-\gamma), h_2\right]} .
    \)
    The surrogate Hamiltonian consists of an Ising term
    \(
    h_1 = \sum_i \ell_i(\vec X),\sigma_z^{(i)} + \sum_{i,j} J_{ij}(\vec X),\sigma_z^{(i)}\sigma_z^{(j)},
    \)
    whose Gibbs state encodes the approximate probability distribution of the neural-network quantum state, and a transverse-field mixer
    \(
    h_2 = \sum_i \sigma_x^{(i)},
    \)
    which induces transitions between configurations. The evolution is implemented via a first-order Trotter decomposition, alternating applications of \(e^{-i h_1 \delta t}\) and \(e^{-i h_2 \delta t}\) for \(N_T\) Trotter steps. Measurements in the computational basis yield the proposed configuration \(\vec v^{(i+1)}\).
    }
\label{fig:Trotter_circuit}
\end{figure}

The surrogate Hamiltonian is written as,
\[
h_1 = \sum_i a_i(\vec X),\sigma_z^{(i)} + \sum_{i<j} J_{ij}(\vec X),\sigma_z^{(i)}\sigma_z^{(j)}
\]
and \(h_2 = \sum_i \sigma_x^{(i)}\) acts as a transverse-field mixer. The corresponding unitary evolution operator is
\[
U(\tau,\gamma) = \exp\left[-i\tau\big(\gamma, h_1 + (1-\gamma), h_2\big)\right],
\]
with \((\gamma,\tau)\in\mathbb{R}\). 
The parameters \(a_i(\vec X)\) and \(J_{ij}(\vec X)\) are learned variational parameters that determine the rotation angles of the single-qubit and two-qubit gates appearing in the quantum circuit implementation of \(U(\tau,\gamma)\). 
The unitary dynamics is implemented through a first-order Trotter decomposition of the time propagator. The resulting circuit (Fig.~\ref{fig:Trotter_circuit}) consists of alternating layers implementing the local-field terms and the pairwise-interaction terms, providing an experimentally feasible approximation of the surrogate Gibbs sampler.

The quantum device prepares an initial product state and applies a sequence of controlled rotations corresponding to the local fields and effective couplings derived from the NQS parameters. After a small number of Trotter steps, measurement of the visible qubits yields samples distributed approximately according to $|\psi_{\theta}(x)|^{2}$. 

\begin{figure}[htbp]
    \centering
    \begin{subfigure}{0.48\linewidth}
        \centering
        \includegraphics[width=\linewidth]{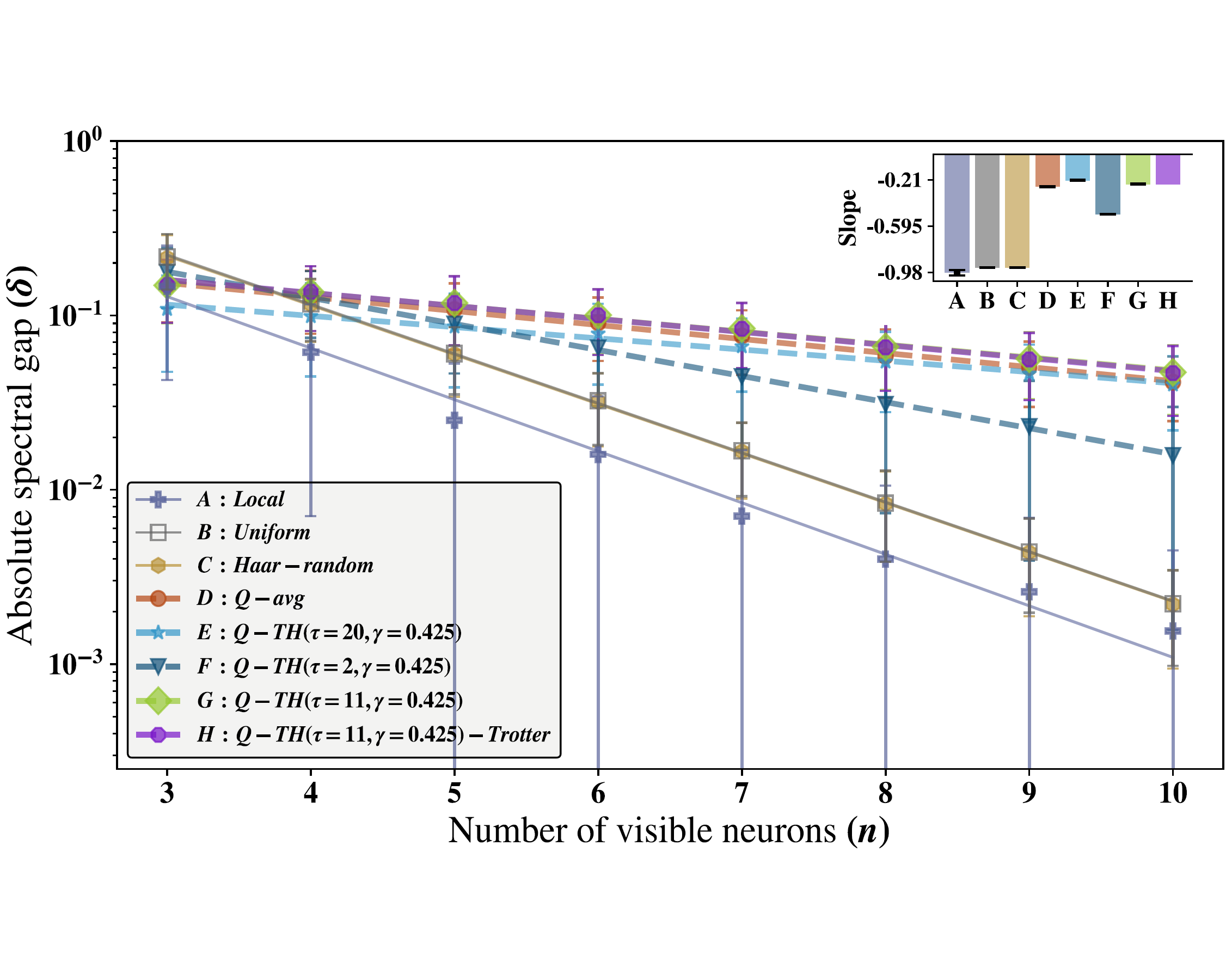}
        \label{fig:Results}
    \end{subfigure}\hfill
    \begin{subfigure}{0.48\linewidth}
        \centering
        \includegraphics[width=\linewidth]{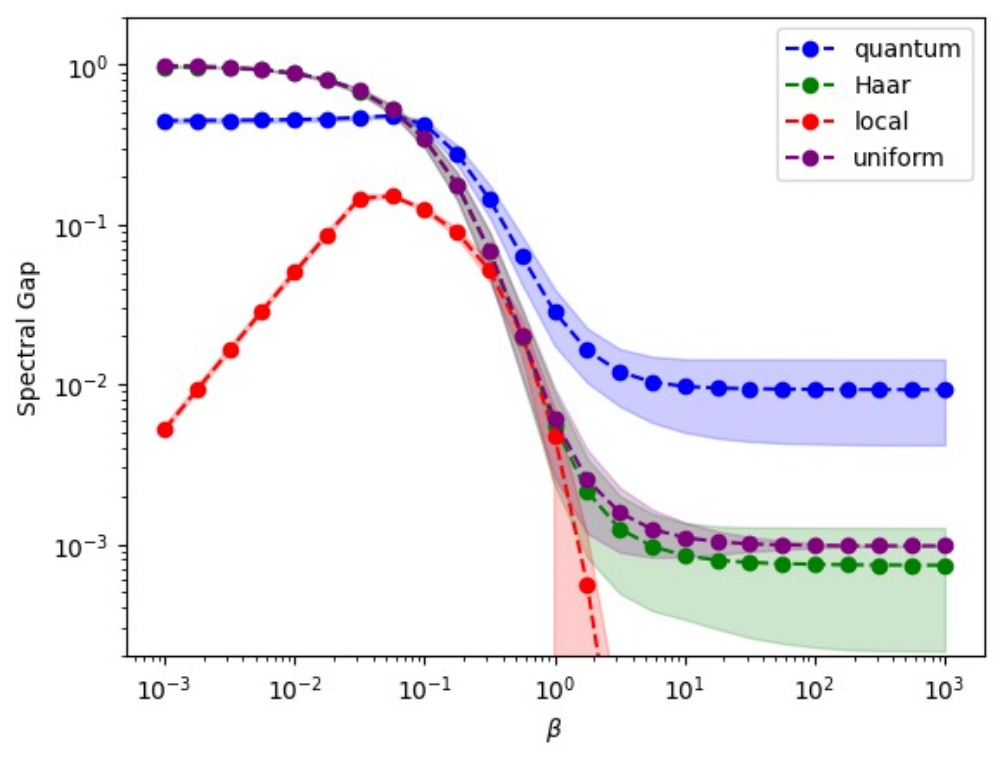}
        \label{fig:Results_Temp}
    \end{subfigure}
    \caption{(left) Absolute spectral gap \(\delta = \lambda_0 - \lambda_1\) of the transition matrix \(T(\vec v^{(i+1)} \mid \vec v^{(i)})\) as a function of the number of visible neurons (n), for different proposal distributions labeled A–H. Classical proposals include (A) local single-spin-flip updates, (B) uniform random proposals, and (C) Haar-random proposals. Quantum-assisted proposals include (D) a quantum-averaged proposal, and (E–G) time-homogeneous (TH) quantum proposals with different evolution times \(\tau\) at fixed mixing parameter \(\gamma\). Case (H) corresponds to the same time-homogeneous quantum proposal implemented using a first-order Trotterized circuit. Inset shows the fitted decay slopes of \(\delta\) versus (n) for each proposal. Quantum proposals (D–H) exhibit a markedly slower decay of the spectral gap compared to classical proposals (A–C), indicating improved mixing time and faster convergence to the stationary distribution. 
    (right) Spectral gap of the transition matrix as a function of inverse temperature \(\beta\) for different proposal distributions. Shaded regions indicate statistical uncertainty from averaging over multiple instances. The quantum proposal maintains a substantially larger spectral gap over a broad range of \(\beta\), particularly in the low-temperature regime, indicating improved mixing properties compared to classical proposals.
    }
    \label{fig:results_method}
\end{figure}

To evaluate the efficiency of this quantum-assisted sampler, we analyzed the spectral gap of the corresponding transition operator, as shown in the numerical experiments of Fig. \ref{fig:results_method}. The quantum sampler maintains a spectral gap that decays with system size at a rate much slower than the classical update MCMC techniques. 
This enhanced mixing time enables the Q-VMC algorithm to avoid long autocorrelation times, thereby producing statistically independent samples. As a direct consequence, gradient estimates and energy evaluations within the VMC loop exhibit significantly reduced variance, resulting in more accurate and stable convergence.

Despite the improved performance, the algorithm preserves polynomial scaling of computational resources - including circuit width, circuit depth, and the number of measurements required for gradient estimation, ensuring that the overall optimization remains efficiently implementable. The circuit width grows linearly with the number of visible units, while the depth is determined primarily by the chosen Trotter step number m. 
Importantly, the Trotterization does not need to accurately approximate the propagator. The sampler succeeds so long as the induced dynamics provide sufficient mixing. The number of samples required depends on the desired accuracy of the gradient estimation. Due to the low correlation among the samples generated from this method, the estimation takes significantly fewer samples compared to classical methods. 

\begin{figure}[htbp]
    \centering
    \begin{subfigure}{0.48\linewidth}
        \centering
        \includegraphics[width=\linewidth]{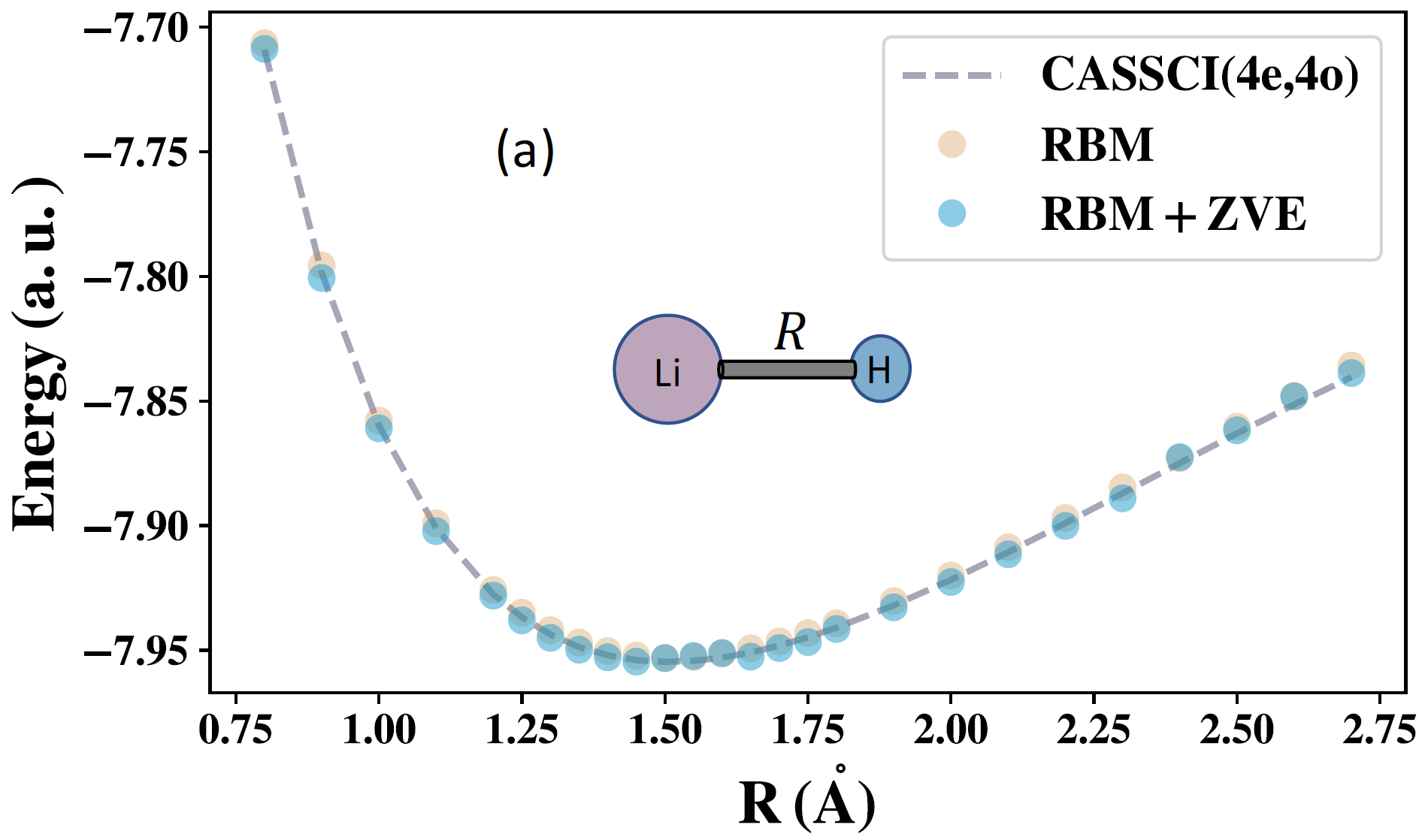}
        \label{fig:LiH}
    \end{subfigure}\hfill
    \begin{subfigure}{0.48\linewidth}
        \centering
        \includegraphics[width=\linewidth]{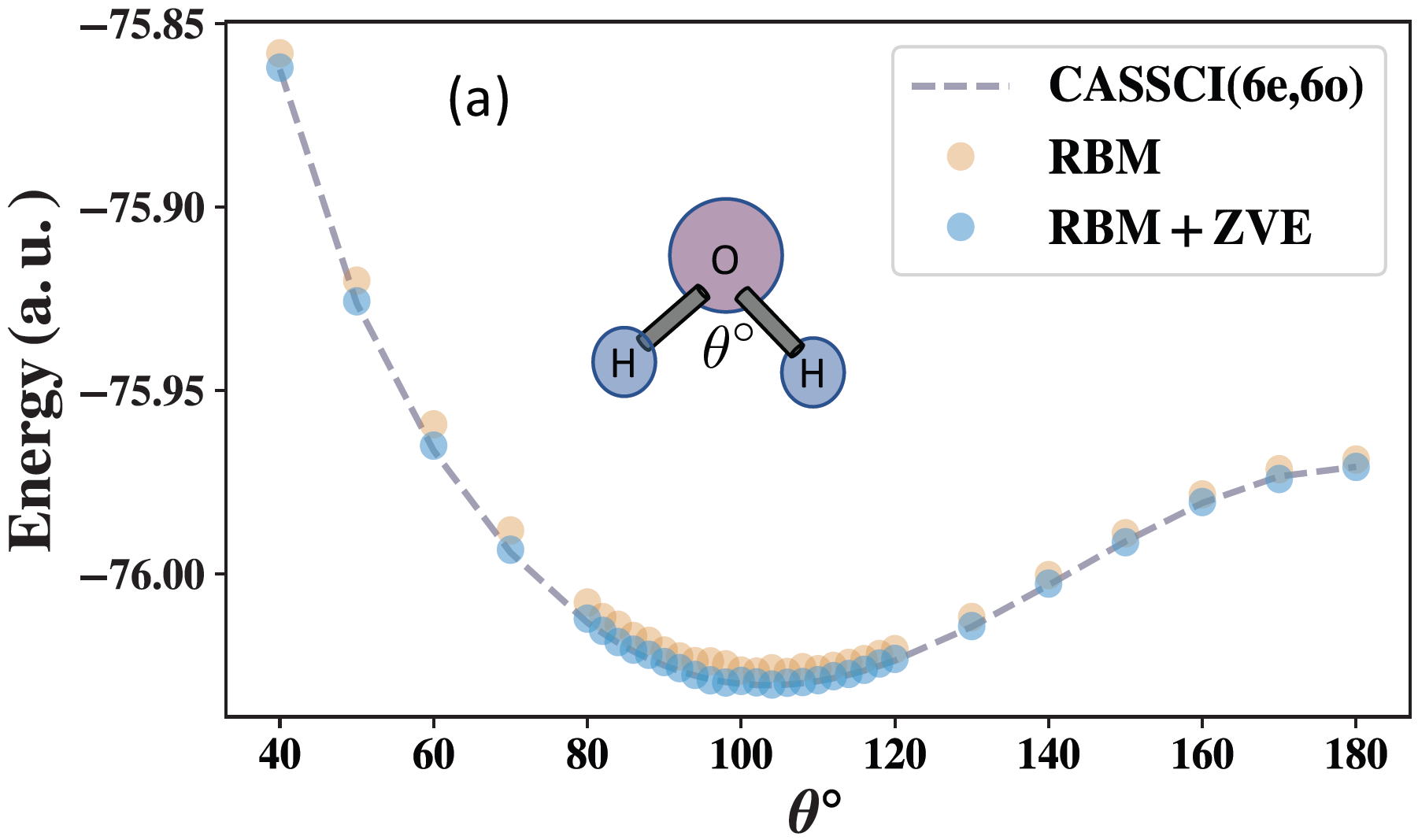}
        \label{fig:H2O}
    \end{subfigure}
    \caption{Benchmarking our algorithm for ground state learning in (a) Li ,  H molecule as a function of distortion in bond length between the Li and H atom (b) H2O molecule as a function of the distortion in bond angle between O and the two H atoms. 
    For computations without ZVE (orange), the minimum energy value from the last few data points in the training protocol is used. For RBM + ZVE (blue), the y-intercept from the Zero variance extrapolation scheme is used. Both methods, especially RBM + ZVE, show good agreement with the exact CASSCI results (dashed gray line) across different bond lengths and bond angles.}
        \label{fig:results_sim}
\end{figure}

We benchmarked the Q-VMC algorithm on representative molecular Hamiltonians and spin models to assess its practical performance. Figure \ref{fig:results_sim} illustrates ground-state learning for the Li–H molecule as a function of bond stretching and for the H$_{2}$O molecule under angular distortion. In both cases, the Q-VMC approach reproduces highly accurate energies when compared to Complete Active Space Configuration Interaction (CASSCI) computations, across a broad range of geometries. 
These benchmarks illustrate how the quantum-assisted sampler maintains stable convergence across the same parameter ranges where standard classical NQS training slows as the geometry moves away from equilibrium.  

The generality of the method allows it to be applied to larger and more complex NQS architectures, including deep or convolutional networks used to represent extended materials and strongly correlated quantum phases. The surrogate Hamiltonian construction accommodates a wide variety of NQS connectivity patterns, making the approach applicable beyond simple RBM-like architectures. In the high-entanglement regime, the improved mixing time manifests as faster and more stable convergence of the variational parameters, as well as more accurate Monte-Carlo estimates of the energy and its gradients, which directly influence the optimization trajectory.

More broadly, the Q-VMC framework provides a blueprint for embedding quantum hardware directly into the optimization loop of NQS-based many-body solvers rather than using quantum circuits solely as variational ansätze or feature maps. This perspective suggests several avenues for future development, including quantum-enhanced training of excited states via orthogonality constraints \cite{Higgott2019VQD, Pfau2020FermiNet}, quantum-assisted preparation of nonequilibrium states for open-system simulations \cite{Yuan2019Theory, Endo2020GeneralProcesses, Chen2024AdaptiveOpen}, and NQS-based generative sampling for molecular and materials design \cite{Carleo2017Solving, Hermann2020PauliNet}. As quantum hardware continues to improve, hybrid training algorithms of this kind may offer a scalable route toward simulating complex physico-chemical systems that lie beyond the reach of conventional classical or purely variational quantum techniques.

\section{Understanding the Learning Dynamics of QML Models}
\label{sec:OTOC}

A central challenge in quantum machine learning (QML) is understanding the structure and navigability of the underlying loss landscape. Trainability issues, such as barren plateaus, highly non-uniform gradients, and sensitivity to initialization, are often closely tied to how information propagates through the model during the optimization process. Out-of-time-order correlators (OTOCs), originally developed to quantify quantum chaos, have recently provided a powerful diagnostic for information spreading in many-body systems. In our recent work \cite{sajjan2023imaginary}, we showed that the imaginary component of the OTOC, a quantity historically overlooked in both physics and QML contexts, encodes detailed geometric information about the learning dynamics of a graph-based neural-network quantum state.

The learner we consider is a restricted Boltzmann machine (RBM) viewed as a bipartite graph $G=(V,E)$ with visible spins $\{v_i\}_{i=1}^n$ and hidden spins $\{h_j\}_{j=1}^p$, each associated with local Pauli operators $\sigma_z(v_i)$ and $\sigma_z(h_j)$. The trainable parameters $X=(a,b,W)$ consist of visible and hidden biases $a\in\mathbb{R}^n$, $b\in\mathbb{R}^p$ and weights $W\in\mathbb{R}^{n\times p}$. As in Ref. \cite{sajjan2023imaginary}, these define an Ising-type Hamiltonian
\begin{equation}
H(X,v,h) = \sum_{i=1}^n a_i \sigma_z(v_i) + \sum_{j=1}^p b_j \sigma_z(h_j)
+ \sum_{i=1}^n \sum_{j=1}^p W_{ij} \sigma_z(v_i)\sigma_z(h_j),
\label{eq:learner_H}
\end{equation}
and a corresponding thermal state
\begin{equation}
\rho_{\mathrm{th}}(X,v,h) = 
\frac{e^{-H(X,v,h)}}{\mathrm{Tr}_{v,h}\left[e^{-H(X,v,h)}\right]}.
\label{eq:thermal_state}
\end{equation}
Sampling $(v,h)$ from $\rho_{\mathrm{th}}$ induces a probability distribution over visible configurations that serves as the amplitude field of the variational ansatz for the driver Hamiltonian.

To probe information transport between visible and hidden units during training, we introduce an OTOC in the Heisenberg picture. For two local operators $U_1(0)$ and $U_2(0)$ evolved under a generator $H_{\mathrm{OTOC}}$,
\begin{equation}
C_{U_1,U_2}(t) = \left\langle U_1^\dagger(0)\,U_2^\dagger(t)\,U_1(0)\,U_2(t)\right\rangle,
\quad
U_2(t) = e^{iH_{\mathrm{OTOC}}t} U_2(0) e^{-iH_{\mathrm{OTOC}}t},
\label{eq:general_otoc}
\end{equation}
with the expectation value taken in $\rho_{\mathrm{th}}(X,v,h)$. In our setting, $H_{\mathrm{OTOC}}$ is chosen to be the learner Hamiltonian in Eq.~(\ref{eq:learner_H}), and the operators are taken to act on a single visible neuron $v_k$ and a single hidden neuron $h_m$. Defining shifted Pauli operators
\begin{equation}
\tilde{\sigma}^\alpha(v_k,0) = \sigma^\alpha(v_k,0) - \kappa_1 \mathbb{I}, \qquad
\tilde{\sigma}^\beta(h_m,0) = \sigma^\beta(h_m,0) - \kappa_2 \mathbb{I},
\end{equation}
with $\alpha,\beta\in\{x,y\}$ and complex shifts $\kappa_{1,2}\in\mathbb{C}$, the OTOC string used in Ref. \cite{sajjan2023imaginary} is
\begin{equation}
C_{\sigma^\alpha,\sigma^\beta}(\kappa_1,\kappa_2,X,t) =
\big\langle \tilde{\sigma}^\alpha(v_k,0)\,\tilde{\sigma}^\beta(h_m,t)\,
\tilde{\sigma}^\alpha(v_k,0)\,\tilde{\sigma}^\beta(h_m,t)\big\rangle_{\rho_{\mathrm{th}}(X,v,h)},
\label{eq:network_otoc}
\end{equation}
where the time evolution of $\sigma^\beta(h_m,t)$ is generated by $H(X,v,h)$.

Theorem II.1 of Ref. \cite{sajjan2023imaginary} shows that, for $\kappa_1=\kappa_2=0$ and $H_{\mathrm{OTOC}}=H(X,v,h)$, the OTOC admits a closed analytic form
\begin{equation}
C_{\sigma^\alpha,\sigma^\beta}(0,0,X,\tau) = 
\cos(\tau) + i \,\big\langle \sigma_z(v_k,0)\,\sigma_z(h_m,0) \big\rangle_{\rho_{\mathrm{th}}(X,v,h)} \,\sin(\tau),
\quad \tau = 4 W_{km} t,
\label{eq:otoc_closed_form}
\end{equation}
and obeys a family of invariants of motion constructed from its real and imaginary parts. Equation~(\ref{eq:otoc_closed_form}) immediately reveals the key role of the imaginary component: the amplitude of
$\mathrm{Im}\,C_{\sigma^\alpha,\sigma^\beta}(0,0,X,\tau)$ is directly proportional to the two-body correlation
$\langle \sigma_z(v_k)\sigma_z(h_m)\rangle_{\rho_{\mathrm{th}}}$ between a visible and a hidden neuron. In phase space, the OTOC trajectories associated with different parameter configurations $X$ lie on circles whose radii are set by this correlator; during training, the learner hops between such trajectories as $X$ is updated, modulating the strength and pattern of correlations between visible and latent units.

To connect this dynamical probe to more conventional correlation measures, we define a covariance-like quantity
\begin{equation}
\eta(X) = \mathrm{Cov}\big(\sigma_z(v_k,0),\sigma_z(h_m,0)\big)_{\rho_{\mathrm{th}}} ,
\end{equation}
and show that it can be expressed solely in terms of OTOCs evaluated at a special time $t_1=\pi/(8|W_{km}|)$ and suitably chosen shifts $(\kappa_1,\kappa_2)$,
\begin{equation}
\eta(X) = C_{\sigma^\alpha,\sigma^\beta}(\kappa_1,0,X,t_1)
+ C_{\sigma^\alpha,\sigma^\beta}(0,\kappa_2,X,t_1)
- C_{\sigma^\alpha,\sigma^\beta}(\kappa_1,\kappa_2,X,t_1),
\label{eq:eta_from_otoc}
\end{equation}
with $\kappa_1=\sqrt{i\langle \sigma_z(v_k,0)\rangle}$ and $\kappa_2=\sqrt{i\langle \sigma_z(h_m,0)\rangle}$, as derived in Eq.~(10) of Ref. \cite{sajjan2023imaginary}. Thus, the imaginary part of the OTOC provides a route to reconstruct not only a two-body correlator but also a covariance that will enter directly into mutual-information bounds.

\begin{figure}[h!]
    \centering
    \includegraphics[width=0.5\linewidth]{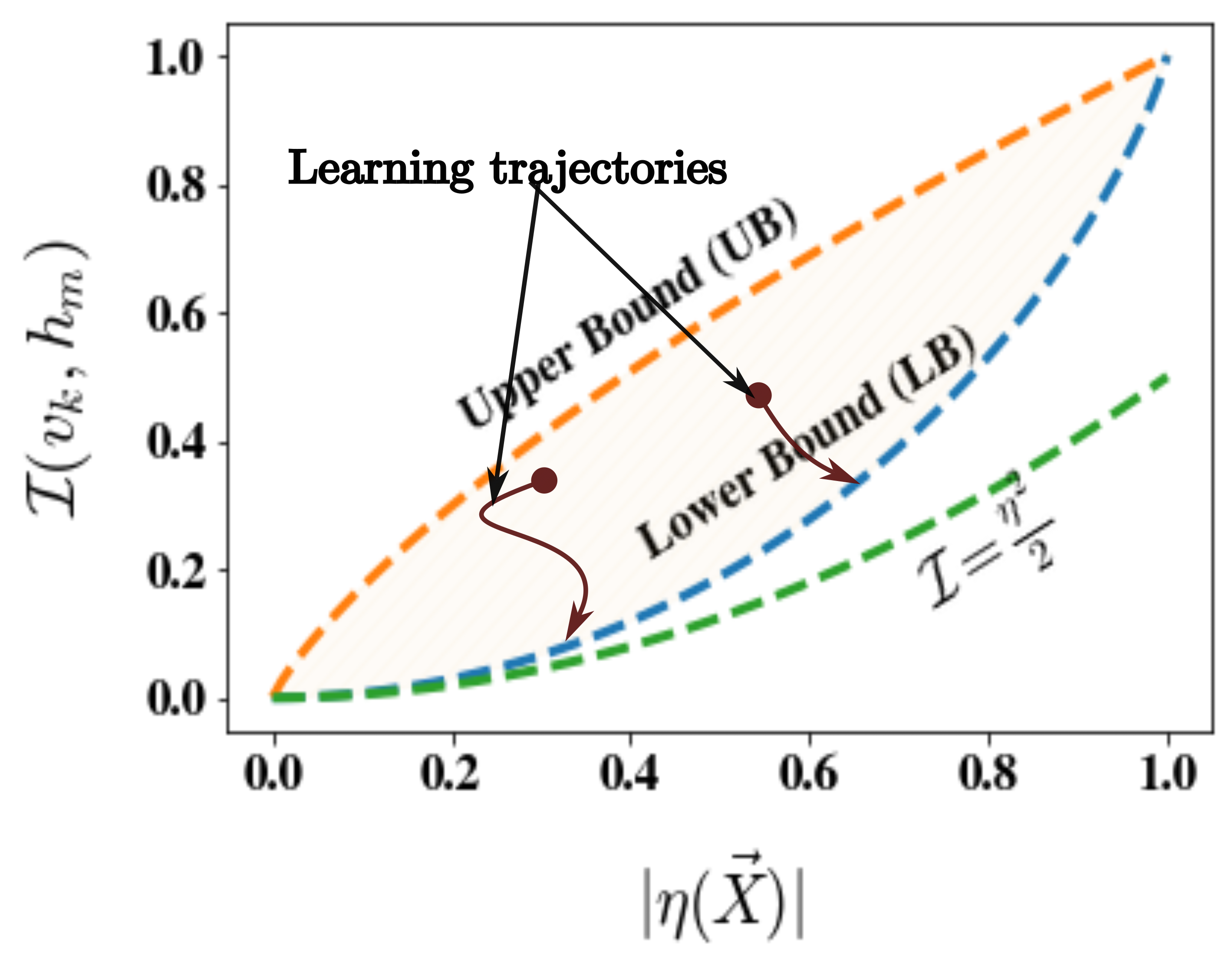}
    \caption{The upper bound(UB) and lower bound (LB) in the convex $I$–$\eta$ space for network G as described in Eqs. \ref{eq:UB_eta}  and \ref{eq:LB_eta} respectively. Provided alongside in green is the conventionally known lower bound for any general bipartition of an arbitrary system \cite{wolf2008area}. The lower bound (LB) for G is thus stricter than the known general bound. Representative learning trajectories of the network G are shown.}
    \label{fig:MI_OTOC_bound}
\end{figure}

The next step is to relate $\eta(X)$ to the quantum mutual information $I(v_k,h_m)$ between the visible neuron $v_k$ and the hidden neuron $h_m$. From the thermal state (\ref{eq:thermal_state}) one can construct the one- and two-spin reduced density matrices $\rho^{(1)}(v_k)$, $\rho^{(1)}(h_m)$, and $\rho^{(2)}(v_k,h_m)$, and hence the mutual information
\begin{equation}
I(v_k,h_m) = S\big(\rho^{(1)}(v_k)\big) + S\big(\rho^{(1)}(h_m)\big) - S\big(\rho^{(2)}(v_k,h_m)\big),
\end{equation}
where $S(\rho) = -\mathrm{Tr}(\rho\ln\rho)$ is the von Neumann entropy. Theorem II.2 of Ref. \cite{sajjan2023imaginary} establishes that, for the RBM learner, $I(v_k,h_m)$ and $\eta(X)$ are constrained to lie inside a convex region of the $(I,\eta)$-plane bounded by
\begin{equation}
\mathrm{LB}(\eta) \le I(v_k,h_m) \le \mathrm{UB}(\eta),
\end{equation}
with an explicit lower bound
\begin{equation}
\mathrm{LB}(\eta) =
2 - 2\,\ell\!\left(\frac{1+\eta/2}{2}\right)
     - 2\,\ell\!\left(\frac{1-\eta/2}{2}\right),
\qquad
\ell(x) = -x\ln x,
\label{eq:LB_eta}
\end{equation}
and a corresponding upper bound 
\begin{equation}
\mathrm{UB}(\eta) =
H \left( \frac{1}{2} + \frac{1}{2}\sqrt{1-|\eta|} \right)
\qquad
H(x) = -x\ln x - (1-x)\ln (1-x) ,
\label{eq:UB_eta}
\end{equation}
Importantly, the lower bound in Eq.~\ref{eq:LB_eta} is strictly tighter than previously known generic bounds of the form $I \ge \eta^2/2$ for arbitrary bipartite systems. The pair $(I(v_k,h_m),\eta(X))$ for every visible–hidden pair thus lives in an emergent $I$–$\eta$ space, and the training dynamics of the learner can be visualized as trajectories within this convex region. Figure~\ref{fig:MI_OTOC_bound} illustrates typical learning trajectories in this space during training of the network, together with the analytic bounds.

To study this emergent space, we train the NQS to approximate the ground state of the driver Hamiltonian $H_{\mathrm{drv}}$ using stochastic reconfiguration (natural gradient) updates of the parameters $X$, following the algorithm of Sorella as implemented in NetKet. At each iteration, the local energy (\(E_{\mathrm{loc}}(v)\))
and derivatives (\(D_i(v)\))
are estimated by Monte Carlo sampling over the distribution $|\psi(v,X)|^2$, where $\psi(v,X)$ is the RBM wave function obtained by marginalizing the thermal state (\ref{eq:thermal_state}) over the hidden spins. The resulting covariance matrix $F_{ij} = \langle D_i^\ast D_j\rangle - \langle D_i^\ast\rangle\langle D_j\rangle$ and force vector $S_i = \langle E_{\mathrm{loc}} D_i^\ast\rangle - \langle E_{\mathrm{loc}}\rangle \langle D_i^\ast\rangle$ define a natural-gradient update
\begin{equation}
X \mapsto X - \ell\,F^{-1}(X)\,S(X),
\end{equation}
with learning rate $\ell$. Because the RBM connectivity is bipartite, each Monte Carlo sweep costs $O(np)$, and computing all parameter updates scales as $O(np N_S)$ for $N_S$ samples per iteration. The construction of all two-spin reduced density matrices, and hence all points in the $I$–$\eta$ space, can be performed using Gibbs sampling from factorizable conditional distributions with overall cost $O(n^2 p^2 N_E)$ for $N_E$ samples. In both stages, the algorithm scales polynomially in the number of visible and hidden neurons.

We benchmark this framework on two families of driver Hamiltonians: the transverse-field Ising model (TFIM) with nearest-neighbor couplings, and a concentric TFIM (c-TFIM) with long-range “concentric” couplings that generate volume-law entangled ground states. Both share the generic form
\begin{equation}
H_{\mathrm{drv}} = -B \sum_{i} \sigma_x(i) - \sum_{i,j} J_{ij} \sigma_z(i)\sigma_z(j),
\end{equation}
with $J_{ij}$ encoding either nearest-neighbor interactions (TFIM) or concentric long-range interactions (c-TFIM), and tuning parameter $g=B/J_0$. For each model, we train RBMs with $n=p=N$ visible and hidden units for system sizes up to $N=24$ spins, and construct the $I$–$\eta$ scatter plot for all visible–hidden pairs at the converged parameter set $X^\ast$.

Several striking features emerge from this analysis. First, for all sizes $N$ and all values of $g$, the trained networks choose representations that lie on the analytic lower bound $\mathrm{LB}(\eta)$ in the $I$–$\eta$ plane. In other words, for a given covariance $\eta(X)$ between a visible and a hidden spin, the RBM spontaneously minimizes the mutual information $I(v_k,h_m)$ within the allowed convex region. This “lower-bound saturation” appears to hold across a wide variety of driver models, suggesting a general learning principle: when tasked with representing a target quantum state, the trained RBM organizes correlations between visible and hidden units to use the minimal mutual information compatible with the required covariance structure.

Second, as the order parameter $g$ is varied through the quantum phase transition of the drivers, the distribution of points in the $I$–$\eta$ space mirrors the behavior of spin correlations in the underlying physical system. For $g\to 0$, where the ground states are strongly correlated, the trained networks exhibit large values of both $\eta(X)$ and $I(v_k,h_m)$. For $g\to\infty$, where the drivers’ ground states become product states, the cloud collapses toward the origin $(I,\eta)=(0,0)$. A well-trained RBM network imbibes the correlation structure of the driver into correlations between visible and hidden neurons during training, even though the learner network has no information about the physical system at the beginning of training.

Third, the emergent $I$–$\eta$ representation distinguishes area-law from volume-law entangled drivers. For the nearest-neighbor TFIM, the spread of points along the lower bound remains relatively narrow for fixed $g$, reflecting a more uniform allocation of correlations across visible–hidden pairs. In contrast, for the c-TFIM with volume-law entangled ground states, the trained networks display a much broader distribution of $(I,\eta)$ values, especially near $g\to 0^+$, indicating many inequivalent ways of assigning correlations between visible and latent units that are compatible with the same learned state. This additional variability can be viewed as a signature of the richer connectivity and entanglement structure of the driver, detected indirectly through the learner.

Taken together, these findings demonstrate that the imaginary part of the OTOC, and the associated $I$–$\eta$ space, provides a powerful dynamical probe of the learning landscape of QML models. They quantify how the RBM “scrambles” information between visible and hidden units during training, how this scrambling is constrained by mutual-information bounds, and how physical features of the driver Hamiltonian, such as phase transitions, correlation ranges, and entanglement scaling, are encoded in the internal geometry of the learner. This perspective suggests that scrambling diagnostics, traditionally used to study quantum chaos and thermalization, can be repurposed as tools for designing and analyzing trainable quantum machine learning architectures for complex quantum systems.

\section{Quantum Machine Learning: From Theory to Real-World Applications}
\label{sec:QML-Bio}

Quantum machine learning (QML) has evolved from a largely theoretical concept rooted in quantum information science into a rapidly developing framework with growing real-world relevance, driven by advances in quantum hardware, algorithms, and data-driven learning. QML is a hybrid computational paradigm that integrates principles of quantum mechanics with classical machine learning techniques to enhance data processing, model expressivity, and computational efficiency \cite{biamonte2017quantum}. By leveraging quantum effects such as superposition and entanglement, QML offers new ways to learn from complex and high-dimensional data.
As a result, QML is now being explored across diverse domains, including drug discovery \cite{bhatia2023quantum}, healthcare \cite{bhatia2023federated, kais2023quantum, bhatia2023handling}, life-science \cite{saggi2024mqml, saggi2026multi, isik2025multimodal}, semiconductor \cite{bhatia2024robustness}, cybersecurity\cite{saggi2023using}, and Finance Portfolio optimization \cite{brandhofer2022benchmarking}. 

\subsection{Quantum-Enhanced Learning Frameworks for Drug Discovery}
\begin{figure}
    \centering
    \includegraphics[width=0.95\linewidth]{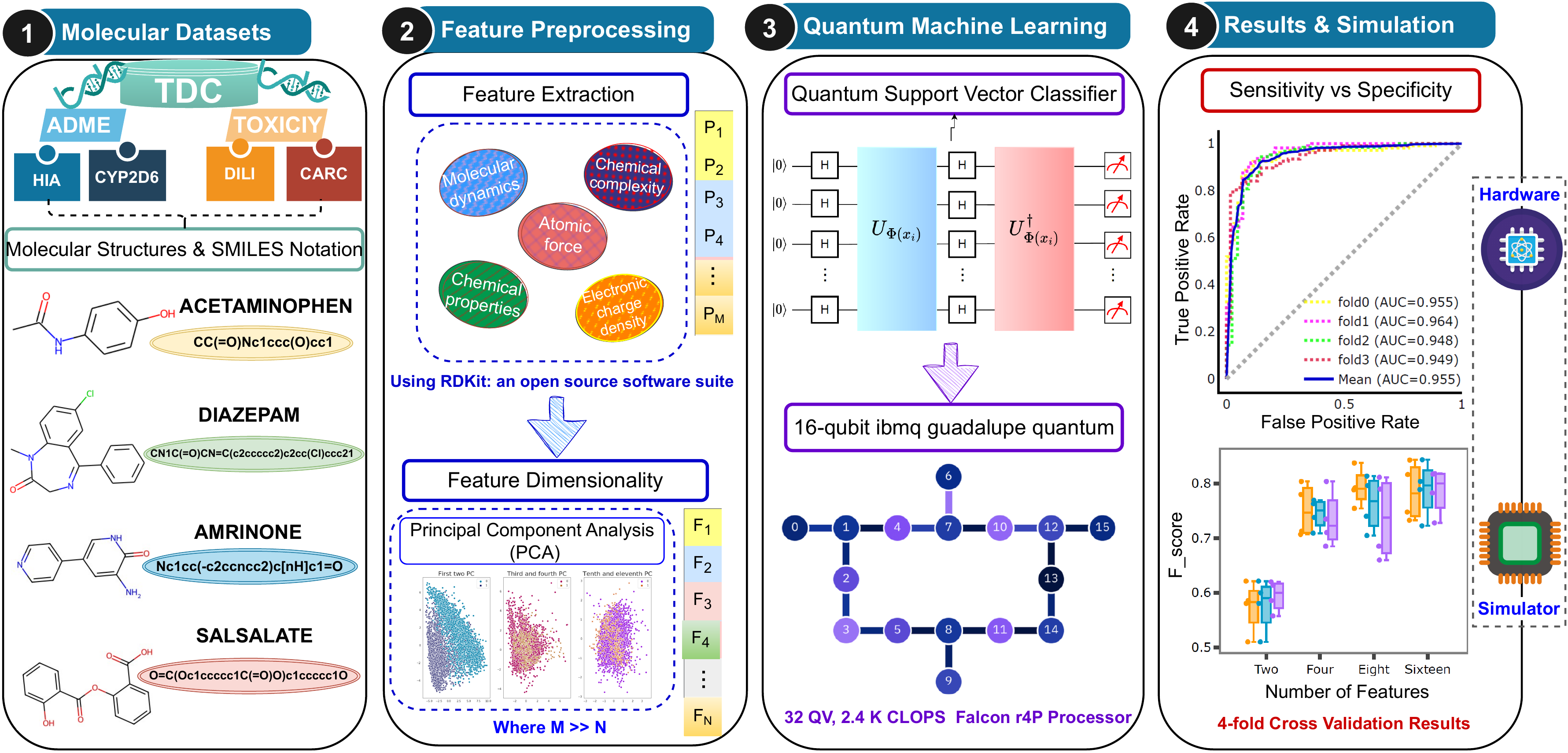}
     \caption{\textbf{An overview of a quantum machine learning framework for predicting ADME and Toxicity properties}. (1) ADME-Tox datasets and their molecular structures and SMILES notations. (2) Feature processing and feature dimension reduction using PCA (3) Quantum classifier for predicting ADME-Tox properties on quantum simulator and hardware (d) Classification metrics.}
     \label{fig:4.1}
\end{figure}

In contemporary drug discovery, accurately evaluating the absorption, distribution, metabolism, and excretion (ADME) profiles, alongside toxicity (Tox) properties of novel chemical entities, remains a significant bottleneck \cite{selick2002emerging}. These assessments are inherently time intensive, costly, and present major challenges for pharmaceutical research and development. Recent advances in computational sciences, including artificial intelligence (AI), big data analytics, and cloud computing, have begun to transform this landscape by enabling predictive models for ADME-Tox properties \cite{maltarollo2015applying}. Emerging at the forefront of this transformation is the integration of quantum computing with machine learning, which has shown promise across chemistry, biomedicine, and engineering. Quantum computational approaches promise to accelerate high-throughput screening of molecular libraries, enabling the evaluation of billions of compounds while substantially reducing both time and cost associated with drug development. Here, we present  a hybrid quantum-classical machine learning framework designed to predict ADME-Tox properties of chemical compounds \cite{bhatia2023quantum}. The  framework integrates a classical support vector classifier with a kernel-based quantum classifier, leveraging the efficiency of quantum kernel methods. Specifically, chemical structures encoded via Simplified Molecular Input Line Entry System (SMILES) strings are transformed into a quantum feature space using string kernels, followed by classification using a quantum-enhanced support vector machine \cite{batra2021quantum}. To assess the performance and practicality of this approach, we conducted extensive simulations on large-scale chemical datasets. Our results demonstrate that the proposed quantum machine learning framework can effectively capture complex molecular features and offers a promising avenue for accelerating ADME-Tox predictions in drug discovery pipelines. Leveraging the potential computational benefits and noise resilience of quantum support vector classifiers, we introduce a quantum machine learning (QML) framework employing QSVC to predict ADME-Tox properties of chemical compounds for drug discovery applications, as depicted in Fig. \ref{fig:4.1} 

\begin{figure}[!ht]
	\centering
	\includegraphics[scale=0.5]{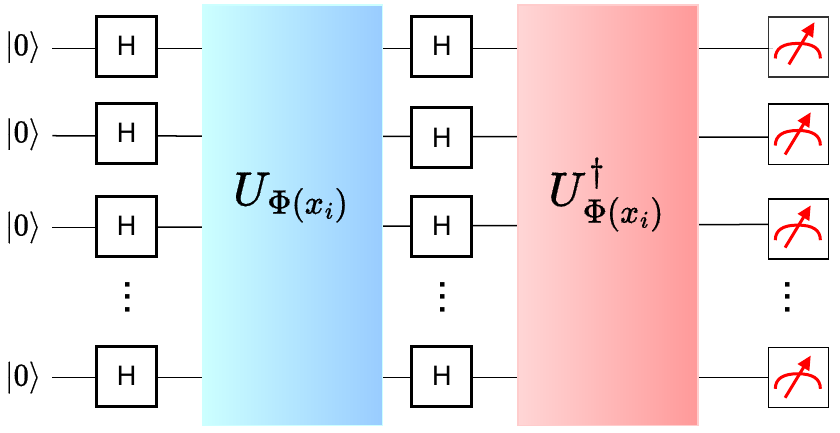}
	\caption{\textbf{A generalized quantum circuit architecture of quantum support vector classifier}. It consists of the encoding circuit, followed by processing (feature map embedding and its inverse) and measurement at the end.}
    \label{fig:4.2}
\end{figure}

The framework consists of: (1) ADME-Tox datasets along with molecular structures and SMILES representations, (2) feature preprocessing and dimensionality reduction using PCA, (3) a quantum classifier implemented on both quantum simulators and hardware to predict ADME-Tox properties using four datasets \cite{huang2021therapeutics}, and (4) evaluation of model performance using classification metrics. In Fig. \ref{fig:4.1}, we present the overall workflow of the proposed framework. To comprehensively investigate the diversity and complexity of ADME-Tox properties, we employed four benchmark chemical compound datasets \cite{huang2021therapeutics}: Human Intestinal Absorption (HIA), CYP2D6 Substrate, Drug-Induced Liver Injury (DILI), and Carcinogenicity. These datasets enable accurate feature learning based on the structural information of drug candidates. All datasets were obtained from the Therapeutic Data Commons (TDC), which provides curated resources for AI/ML tasks in multiple therapeutic modalities, including small molecules, macromolecules, and cell and gene therapies to support effective drug development and improved patient safety throughout the lifecycle of the medicine.
The support vector classifier (SVC) is a well-known classical machine learning algorithm, also called the `kernel' method, for classification and regression. Suppose that given \textit{N} training data points ($x_i$) and associated labels $y_i \in \{-1,1\}$, which are spelled in \textit{dimensions D} , where $1 \leq i \leq N$. The objective is to find a hyperplane that maximizes the margin between the classes in space. In SVC, it computes the inner product between pairs of datapoints $\braket{x_i|x_j} \in \mathbb{R}$. If the data points cannot be separated linearly, then transform the input data points to a high-dimensional feature space ($\varphi$) via a mapping function and determine a hyperplane with maximum margin. The technique `kernel trick' in SVC enables handling nonlinear datapoints, which relies on the inner product between pairs of mapping kernel functions such that $K(x_i, x_j)=\varphi(x_i).\varphi(x_j)$. Thus, the optimization objective function of SVC with the kernel trick is given as:
\begin{equation}
\begin{split}
\text{max} ~f(c_1, ..., c_N)& =\sum_{i=1}^{N}c_i-\frac{1}{2} \sum_{i=1}^{N} \sum_{j=1}^{N} y_i c_i (\varphi(x_i).\varphi(x_j))y_jc_j\\
& = \sum_{i=1}^{N}c_i-\frac{1}{2} \sum_{i=1}^{N} \sum_{j=1}^{N} y_i c_i K(x_i.x_j)y_jc_j
\end{split}
\end{equation}
subject to $\sum_{i=1}^{N}c_i y_i=0$, and $0\leq c_i \leq 1/2N\lambda$, where $\lambda$ controls the hard margin classifier. Although there exist various kernel functions, such as linear, polynomial, Gaussian (RBF), and sigmoid, to solve different problems.
Kernel-based learning methods have been successfully applied to problems involving high-dimensional input spaces. 
The quantum support vector classifier (QSVC) \cite{2025arXiv250517756S} is an extension of the classical support vector classifier (SVC) that leverages a quantum feature space to enhance computational power. The quantum feature space $\ket{\varphi(x_i} \bra{\varphi(x_j)}$ is a high-dimensional complex-valued Hilbert space, which allows for the exploitation of quantum computing capabilities. An architecture of QSVC is shown in Fig. \ref{fig:4.2}. Here, a quantum kernel function is employed to map the data into a quantum feature space represented as
\begin{equation}
    K(x_i, x_j)=|\braket{\varphi(x_i}|{\varphi(x_j)}|^{2}
\end{equation} 

where $\varphi(x_i)$ denotes the first step is to encode classical points $x \in \mathbb{R}^n$ into a quantum data using a \textit{n}-qubit parameterized quantum circuit (PQC) $\ket{\varphi(x)}=U(x)\ket{0^n}$, where \textit{U} is a unitary operation. We used the ZZFeaturemap with a linear entanglement layer (depth=2). Although we can vary the depth of the feature map circuit to introduce more entanglement and repeat($\times l$) the encoding circuit. The number of qubits depends on the number of features outlined in our dataset. The quantum kernel is estimated by evolving a reference state $\ket{0}^n$ as
\begin{equation}
 K(x_i, x_j)=| \bra{0^n}U^{\dagger}(x_i) U(x_i) \ket{0^n}|^{2}
\end{equation}

\begin{figure}
    \centering
    \includegraphics[width=1.0\linewidth]{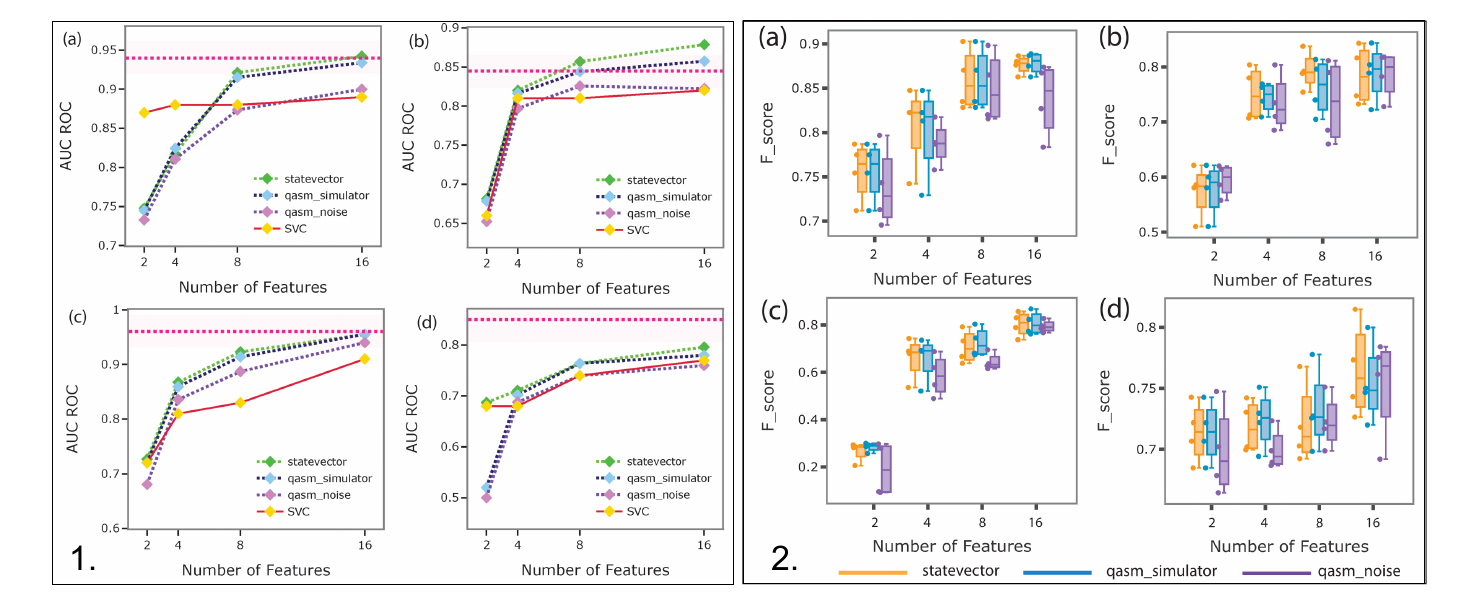}
    \caption{\textbf{1.: The mean AUC ROC curves of the quantum classifier across all quantum simulators}. Here, we reported the mean AUC ROC values for different feature sets used to train the QSVC model. Fig (a-d) shows the CARC, DILI, HIA, and CYP2D6 datasets with the PCA feature reduction method. The statevector backend is in green dotted line, (qasm simulator (QA) in blue), and (a qasm simulation with noise (QN), represented with purple dotted line). The classical SVC is denoted by a solid line (red), and the shaded line is for the SVC considering 2048-bit RDKit fingerprints.  2. \textbf{Graphical representation of boxplots for F-score performance metric of quantum simulators (statevector as orange), (qasm simulator in blue), and (qasm with noise as purple)}. Here, we applied 4-fold cross-validation across all simulations and presented the F-scores for all datasets (CARC, DILI, HIA, and CYP2D6) with different numbers of features for all quantum simulators in (a-d), respectively. In boxplots, the median value is denoted by the horizontal line, and the box defines upper and lower quantiles. The data points are the statistical outliers.}
     \label{fig:4.3}
\end{figure}

\begin{table*}[!ht]
\caption{Classification performance metrics of QSVC considering 16 features on QASM noise model.}
\label{tab:qsvc_metrics}
\centering
\begin{tabular*}{\textwidth}{@{\extracolsep{\fill}}lcccccccc}
\toprule
 & \multicolumn{4}{c}{ADME Dataset: HIA} & \multicolumn{4}{c}{ADME Dataset: CYP2D6} \\
\cmidrule(lr){2-5} \cmidrule(lr){6-9}
Metrics & Fold 1 & Fold 2 & Fold 3 & Fold 4 & Fold 1 & Fold 2 & Fold 3 & Fold 4 \\
\midrule
Accuracy  & 0.863 & 0.881 & 0.881 & 0.863 & 0.706 & 0.664 & 0.746 & 0.735 \\
Precision & 0.703 & 0.750 & 0.774 & 0.777 & 0.842 & 0.730 & 0.847 & 0.752 \\
Recall    & 0.918 & 0.785 & 0.889 & 0.792 & 0.694 & 0.657 & 0.729 & 0.800 \\
F-score   & 0.796 & 0.767 & 0.827 & 0.785 & 0.761 & 0.691 & 0.784 & 0.775 \\
\midrule
 & \multicolumn{4}{c}{Toxicity Dataset: DILI} & \multicolumn{4}{c}{Toxicity Dataset: CARC} \\
\cmidrule(lr){2-5} \cmidrule(lr){6-9}
Metrics & Fold 1 & Fold 2 & Fold 3 & Fold 4 & Fold 1 & Fold 2 & Fold 3 & Fold 4 \\
\midrule
Accuracy  & 0.757 & 0.841 & 0.766 & 0.849 & 0.821 & 0.785 & 0.845 & 0.747 \\
Precision & 0.796 & 0.862 & 0.775 & 0.872 & 0.907 & 0.895 & 0.900 & 0.844 \\
Recall    & 0.741 & 0.814 & 0.789 & 0.842 & 0.830 & 0.767 & 0.849 & 0.730 \\
F-score   & 0.727 & 0.818 & 0.782 & 0.817 & 0.867 & 0.826 & 0.873 & 0.783 \\
\bottomrule
\end{tabular*}
\end{table*}
The quantum circuit with linear entanglement for 8-dimensional classical input features is given in Fig. \ref{fig:4.3}, where \textit{H} denotes the Hadamard gate to create superposition, and $P(\theta)$ is a phase gate. A unitary operation (\textit{U}) is applied to quantum registers initialized in the $\ket{0}$ state, and then the inverse operation \textit{U}$^\dagger (x)$ is applied. Finally, a measurement operator is applied to estimate the quantity of interest (i.e., the target variable). To showcase the potential advantage of a quantum machine learning algorithm, the proposed framework conducted experiments in two phases: In the initial phase, the experiments were executed in a noiseless quantum environment with the python-based Qiskit's statevector simulator (SV), and qasm simulator (QA), which means that the simulations assumed perfect conditions without any form of noise or errors \cite{robertson2023squeeze}. The QSVC model was evaluated on balanced HIA, CYP2D6, DILI, and CARC datasets, showing that performance consistently improves as feature dimensionality increases from 2 to 16, while remaining compatible with current quantum hardware limits. QSVC achieved strong predictive capability for HIA and CARC (AUC $\approx$ 0.93–0.95) and stable performance for DILI (AUC $\approx$ 0.85–0.87), with robustness even under noise. CYP2D6 remained the most challenging task, performing moderately (AUC $\approx$ 0.77–0.79) and below a fully featured classical SVC, indicating that some molecular properties may require higher-dimensional embeddings. Compared with classical models, QSVC often matched or exceeded performance using far fewer features, remained consistent on real IBM hardware, and demonstrated reliable generalization as shown in Fig. \ref{fig:4.3}. Overall, the results highlight that QSVC effectively captures complex molecular relationships, delivers competitive ADME-Tox prediction with reduced feature requirements, and shows promising signs of emerging quantum advantage as hardware capability grows.

\subsection{Quantum-Driven Approaches in Cancer Biology and Biomarker Discovery}

Quantum machine learning offers novel opportunities to resolve, accelerate, and refine computational biology challenges. In \cite{saggi2024mqml}, our study aims to investigate the performance of multi-omic data integration and high-dimensional feature encoding for cancer prediction (primary tumor or normal solid tissue) and to assess feature importance using Quantum-Classical Machine Learning. Moreover, in \cite{saggi2026multi}, we focus on advancing QML for biomedical applications, particularly in multi-omic data integration for lung cancer subtype classification. Traditional machine learning approaches often struggle with the high dimensionality and heterogeneity of omics data. To address this, we developed a Multi-omic Quantum Machine Learning Lung Subtype Classification (MQML-LungSC) framework that leverages Quantum Neural Networks (QNNs) for enhanced classification performance. By encoding multi-omic features, including DNA methylation, RNA-seq, and miRNA-seq, our approach identifies key molecular markers distinguishing lung adenocarcinoma (LUAD) from lung squamous cell carcinoma (LUSC). Our results demonstrate that QNNs, particularly with 256 encoded features, outperform classical machine learning models such as Logistic Regression (LR), Multi-Layer Perceptron (MLP), and Support Vector Machines (SVM). The QNN256 model achieved training and test accuracies of 0.95 and 0.90, respectively, with AUCs of 0.99 and 0.96, respectively. Additionally, our methodology enables the identification of top-hit biomarkers using model weights, facilitating biomarker discovery for potential clinical applications. This study underscores the promise of quantum-classical hybrid models in personalised medicine and sets the stage for future advancements in QML-driven cancer diagnostics. 

\begin{figure}
    \centering
    \includegraphics[width=0.95\linewidth]{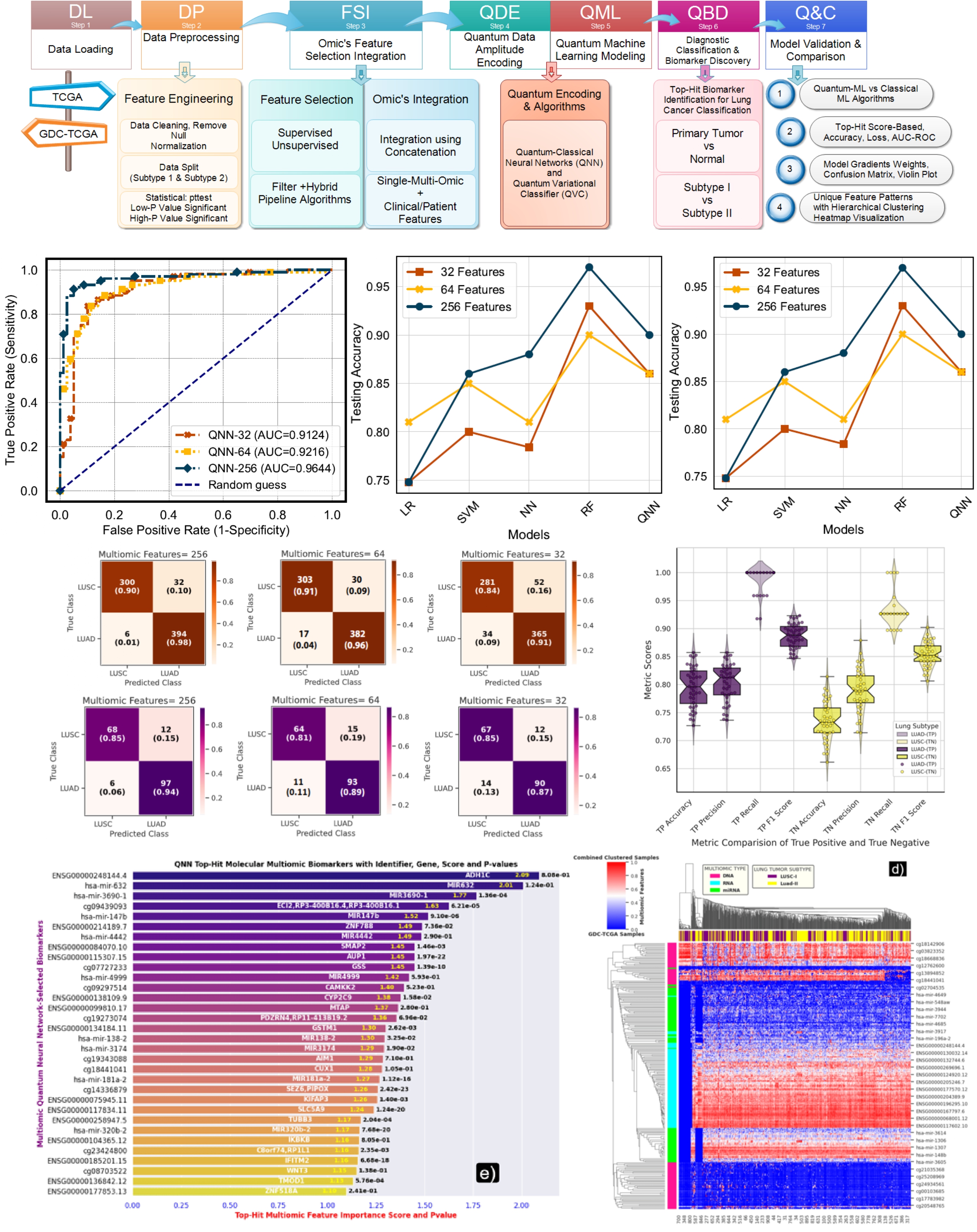}
    \caption{Workflow of multi-omics integration and classification using quantum neural networks.}  
     \label{fig:4.4}
\end{figure}

Fig. \ref{fig:4.4} shows the workflow of multi-omics integration and classification using quantum neural networks. \textbf{(a)} Schematic representation of the overall pipeline of the MQML-QNN framework. \textbf{(b)} Data acquisition from GDC-TCGA, including (i) DNAme, (ii) RNA-seq, (iii) miRNA-seq, and (iv) clinical and survival attributes of patients.\textbf{ (c)} Data preprocessing and feature engineering using t-test p-values for each omic data type to differentiate between subtypes I and II. \textbf{(d)}-\textbf{(e)} Feature selection process involving: (i) Four feature selection models: RF, MI, PCA, and Chi. (ii) AUC-ROC analysis, (iii) Hierarchical clustering based on sample vs. feature, and pairwise feature similarity using distance metrics. \textbf{(f)} Integration of multi-omic data resulting in 256 features from 915 common patients. \textbf{(g)} Quantum amplitude encoding with three feature dimensions: 32, 64, and 256, and the quantum circuit with a dense layer for diagnostic classification of LUAD versus LUSC lung datasets. \textbf{(h)} Visualization of features through violin plots, confusion matrices, and heatmaps.

\subsection{Quantum-Driven Agro-Climate Modeling}

In recent years, quantum computing has emerged as a powerful tool in agriculture, particularly for processing highly correlated, high-dimensional climate data \cite{yadav2025knowledge}. Matrix Product State (MPS) quantum classifiers \cite{bhatia2018quantifying}, a type of one-dimensional tensor network, have been shown to efficiently encode classical agricultural data, such as temperature, humidity, wind speed, solar radiation, sunshine hours, and evapotranspiration (ET$_{o}$) \cite{saggi2019reference, saggi2022survey}, into quantum-entangled states for classification tasks.
MPS constrains the extent of entanglement through its bond dimensions. It is a tensor network method in which tensors are connected in a one-dimensional geometry. In principle, any pure quantum state can be represented by appropriately decomposing its coefficients. In the MPS formalism, a pure quantum state $|\phi\rangle$ is expressed as
\begin{equation}
|\phi\rangle =
\sum_{\sigma_1,\sigma_2,\ldots,\sigma_L}
\mathrm{Tr}\!\left[
M^{\sigma_1}_1
M^{\sigma_2}_2
\cdots
M^{\sigma_L}_L
\right]
|\sigma_1,\sigma_2,\ldots,\sigma_L\rangle ,
\label{eq:4_1}
\end{equation}
where $M^{\sigma_i}_i$ are complex square matrices, $d$ denotes the local dimension, $\sigma_i$ represents the local basis indices (i.e., $\{0,1\}$ for qubits), and $\mathrm{Tr}(\cdot)$ denotes the trace of the matrices. In quantum mechanics, the $N$ independent systems can be combined by performing the tensor product operation on their respective state vectors~\cite{stoudenmire2016supervised}. The feature map is considered as shown in Eq.~\eqref{eq:feature_map}:
\begin{equation}
\label{eq:feature_map}
\phi_d(x) = \phi_{s_1}(x_1) \otimes \phi_{s_2}(x_2) \otimes \cdots \otimes \phi_{s_N}(x_N),
\end{equation}
where $s_j$ are indices running over the local dimension $d$ such that $d = \{s_1, s_2, \ldots, s_N\}$. Therefore, each state vector $x_j$ is mapped to the full feature map $\phi(x)$ in a $d$-dimensional space. Fig.~4.2 shows the tensor diagram of the full feature map $\phi(x)$. Before illustrating the MPS tensor network, it is crucial to encode a classical machine learning dataset into a quantum state. Consider a classical dataset $S = \{(x^d, y^d)\}_{d=1}^{D}$ for binary classification, where $y^d \in \{0, 1\}$ are class labels for $N$-dimensional input vectors such that $x^d \in \mathbb{R}^N$. The input vectors are normalized to lie in $[-\pi, \pi]$. Thus, the qubit $\phi$ is represented as in Eqs.~\eqref{eq:qubit_single} and~\eqref{eq:qubit_full}:
\begin{equation}
\label{eq:qubit_single}
\phi^d_n = \cos(x^d_n) \, |0\rangle + \sin(x^d_n) \, |1\rangle,
\end{equation}
\begin{equation}
\label{eq:qubit_full}
\phi^d_n =
\begin{bmatrix}
\cos(x^d_1) \\ \sin(x^d_1)
\end{bmatrix}
\otimes
\begin{bmatrix}
\cos(x^d_2) \\ \sin(x^d_2)
\end{bmatrix}
\otimes \cdots \otimes
\begin{bmatrix}
\cos(x^d_N) \\ \sin(x^d_N)
\end{bmatrix}.
\end{equation}

The $ N$-dimensional input vectors $x^d \in \mathbb{R}^N$ are mapped to a product state on $N$ qubits by using the feature map as shown in Eq.~\eqref{eq:feature_map}. The full quantum data is represented as the tensor product $\phi^d = \bigotimes_{n=1}^{N}\phi^d_n$~\cite{huggins2019towards}. Thus, the preparation of the quantum state is efficient as it only needs single-qubit rotations to encode each segment of the classical dataset $n = \{1, 2, \ldots, N\}$ in the amplitude of a qubit, with no high cost for such encoding.  Similar to a classical dataset for binary classification, a quantum dataset for binary classification is denoted as a set: $S_q = \{(\phi^d, y^d)\}_{d=1}^{D}$  where $y^d \in \{0, 1\}$ are class labels for $2^N$-dimensional input vectors such that $\phi^d \in \mathbb{C}^{2^N}$. It can be easily verified that the quantum data, as the output of a quantum circuit, is in a superposition state. To assess the quality of the actual and predicted values in the dataset, a cost function is needed. It measures the difference between actual and predicted values and is given by Eq.~\eqref{eq:cost_function}:
\begin{equation}
\label{eq:cost_function}
J_\theta = \frac{1}{D} \sum_{d=1}^{D} \left( M_\theta(x^d) - y^d \right)^2,
\end{equation}
where $x^d$ and $y^d$ are the input and class labels respectively, $M$ is a qubit operator, $\theta$ represents the set of parameters defining the unitaries, and $D$ is the total number of data points. This calculates the average deviation of the model’s predictions from the actual values. 
The goal is to minimize the cost function, i.e., it must be close to zero. 
In this context, historical meteorological data were preprocessed, normalized, and categorized into binary classes of ET$_{o}$ (LOW, MEDIUM, HIGH) to form distinct classification tasks \cite{bhatia2019matrix}. The MPS quantum circuit, composed of single-qubit rotations, CNOT gates, and ancilla-assisted nonlinear unitaries, was trained on 80\% of the data and tested on the remaining 20\%, achieving accuracies of 73–80\% across samples, along with high sensitivity and specificity. 
\begin{figure}
    \centering
    \includegraphics[width=0.85\linewidth]{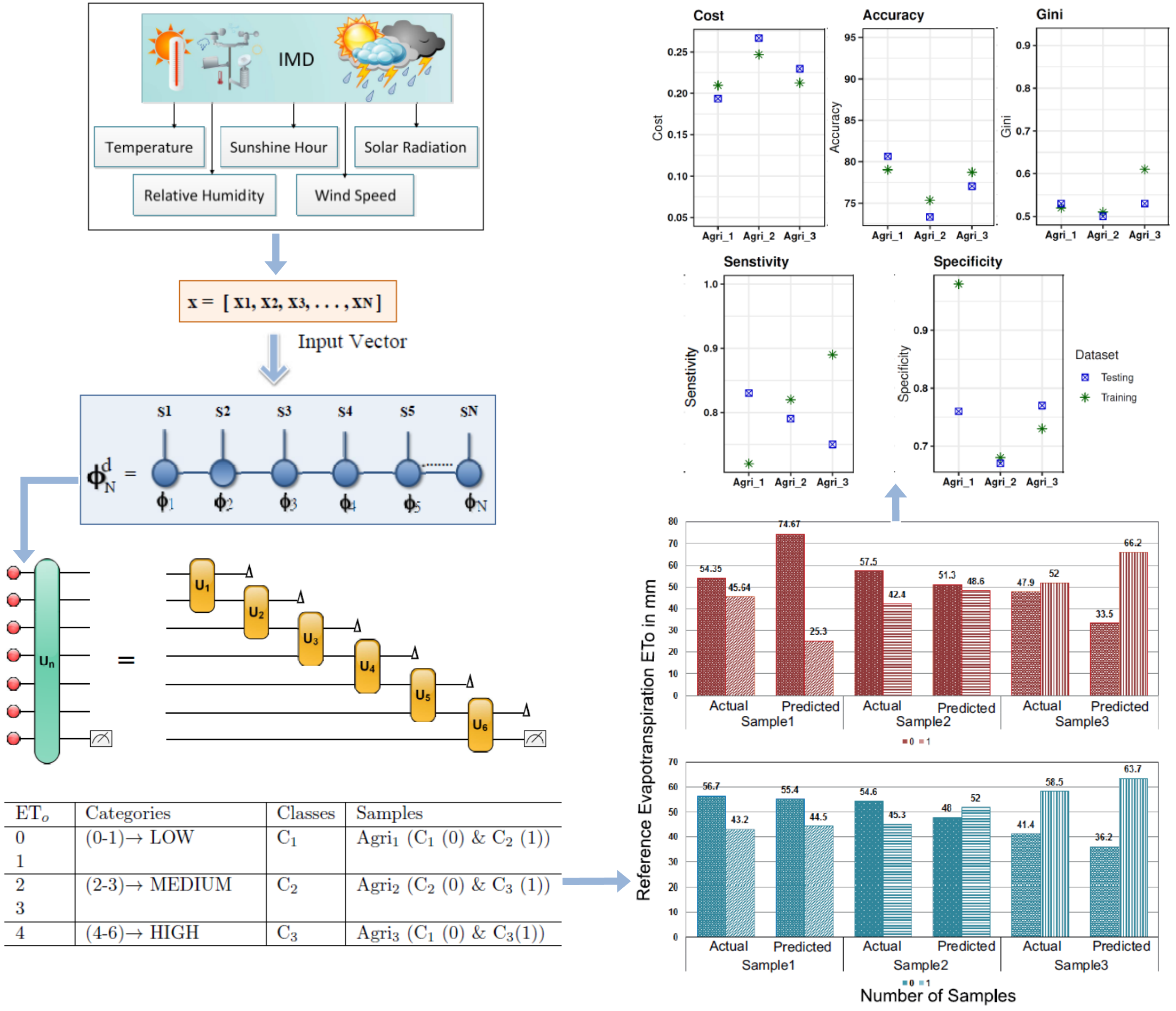}
    \caption{Schematic representation of the proposed Matrix Product State (MPS)–based quantum classification framework for the IMD weather dataset,  including pairwise partitioning of the dataset, normalization, binary classification setup, encoding of classical input vectors into quantum states, construction of the MPS tensor network, measurement, validation, and final classification output for low, medium, and high Reference Evapotranspiration ET$_{o}$ categories.}  
     \label{fig:4.5}
\end{figure}
Fig. \ref{fig:4.5} illustrates the overall architecture and processing steps of the proposed MPS quantum classifier applied to the Indian Meteorological Department (IMD), Pune weather dataset. Daily meteorological data for Patiala in 2014, 2015, and 2016 have been used. It consists of the following parameters: maximum and minimum air temperature ($T_{\text{max}}$, $T_{\text{min}}$), relative humidity (RH), wind speed ($u_2$), solar radiation ($R_s$), sunshine hours ($I_s$), and evapotranspiration (ET$_{o}$). Initially, the weather dataset is partitioned pairwise into three samples: Agri1 (Sample1), Agri2 (Sample2), and Agri3 (Sample3). The dataset is then normalized to ensure uniform feature scaling before applying binary classification, where the input feature vectors $X_1, X_2, \ldots, X_n$ and corresponding target labels $Y_1, Y_2, \ldots, Y_n$ represent low, medium, and high categories for the estimation of reference evapotranspiration (ET$_{o}$). The combined pairwise samples are further divided into training and test subsets. The training and testing datasets are mapped into a tensor network state using Eq.~(4.2). Each classical input vector is encoded into a quantum state using Eq.~(4.3), enabling the representation of classical weather parameters in a quantum Hilbert space. Subsequently, the tensor product of each input quantum state is taken to form complete quantum data, as defined in Eq.~(4.4), which is suitable for processing within a quantum circuit. This representation corresponds to a Matrix Product State with multiple sites. The encoded quantum states are then processed by the MPS quantum classifier, which performs tensor contractions and measurements to obtain classification results. Finally, the validation process evaluates the model's performance on the test dataset, yielding classification outputs for the low, medium, and high ET$_{o}$ categories. The method efficiently captures correlations among climatic variables, mitigates overfitting through adaptive bond dimensions, and requires a small number of qubits, making it suitable for near-term quantum hardware.  Visualization confirmed a strong correspondence between predicted and actual ET$_{o}$ values. The advantages of this approach include efficient handling of large datasets, robust learning capability, and potential scalability to other agricultural parameters. Practical applications include precision irrigation planning, climate-adaptive farming, and integration into smart agriculture systems for real-time decision-making \cite{saggi2022survey}. Overall, this work demonstrates that MPS quantum classifiers provide a promising framework for leveraging quantum computing to enhance agricultural data analysis and improve resource management.

\section{Federated Quantum Machine Learning: Distributed Intelligence in the Quantum Era.}
\label{sec:FQML}

Quantum federated learning (QFL) is an emerging paradigm that integrates the principles of federated learning with quantum information processing to enable collaborative model training without centralized data sharing. In classical federated learning, multiple clients train local models on their private data and exchange only model updates with a central server, preserving data privacy and reducing the transmission of raw data. QFL extends this idea to quantum settings, where clients may possess quantum data, quantum models (such as variational quantum circuits), or hybrid quantum–classical architectures. By allowing quantum nodes to train local quantum models and share parameters or measurement statistics instead of quantum states or raw data, QFL aims to combine privacy preservation with the expressive power of quantum machine learning, particularly in scenarios involving distributed, sensitive, or heterogeneous datasets \cite{chen2021federated}.

QFL introduces uniquely quantum challenges and opportunities, including quantum noise, limited qubit counts, and client heterogeneity in quantum hardware. Advanced QFL frameworks explore techniques such as layer-wise or parameter-wise aggregation of variational quantum circuits, quantum-aware optimization strategies, and noise-resilient protocols that account for non-IID data and device-dependent errors. Recent research also investigates secure aggregation, quantum communication advantages, and hybrid schemes where quantum clients collaborate with classical servers. As quantum hardware matures, QFL is expected to play a key role in scalable, privacy-preserving quantum intelligence, enabling distributed learning across quantum networks, cloud-based quantum processors, and future quantum internet infrastructures.

  \begin{figure*}[!ht]

\includegraphics[scale=0.28]{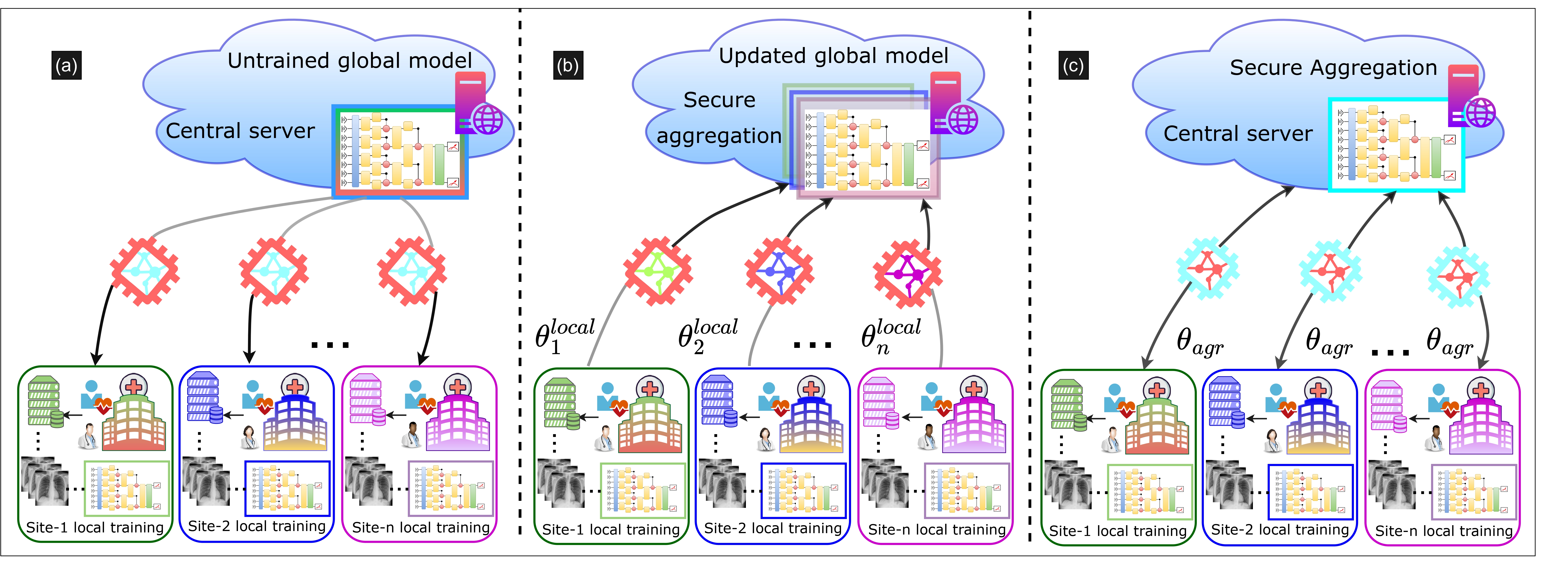}
\caption{Illustration of the Quantum Federated Learning (QFL) framework, showcasing the key steps involved in the collaborative training of quantum models across multiple clients. (a) An untrained global model is sent to all clients/hospitals. (b) Each client independently trains its local quantum circuit using private data, subsequently sharing only the updated quantum parameters with a central aggregator \cite{137}. (c) The global model is refined through aggregation techniques, ensuring data privacy while leveraging quantum computing to enhance learning efficiency and performance. }
 \label{qfl}
\end{figure*}

Fig.~\ref{qfl} presents the overall workflow of the proposed QFL framework, which enables collaborative training of quantum models while preserving data privacy across distributed healthcare institutions. The process begins by initializing an untrained global quantum model on a central server, which is then broadcast to all participating clients or hospitals. Each client receives an identical copy of the global model and performs local training using its private dataset, ensuring that sensitive patient data remains on-site.

During local optimization, clients update the parameters of their quantum circuits independently and transmit only the learned quantum parameters to the central aggregator, rather than sharing raw data or intermediate activations. This parameter-based communication mechanism significantly reduces privacy risks while enabling collaborative learning. Upon receiving updates from all participating clients, the central server aggregates the quantum parameters using federated aggregation strategies to construct an improved global model. This iterative exchange between clients and the server continues until convergence, allowing the framework to exploit the advantages of quantum computation for enhanced learning efficiency and performance in a privacy-preserving federated setting~\cite{137}.

\subsection{Federated quantum machine learning and its applications}

 \begin{figure*}[t]
	
\includegraphics[scale=0.42]{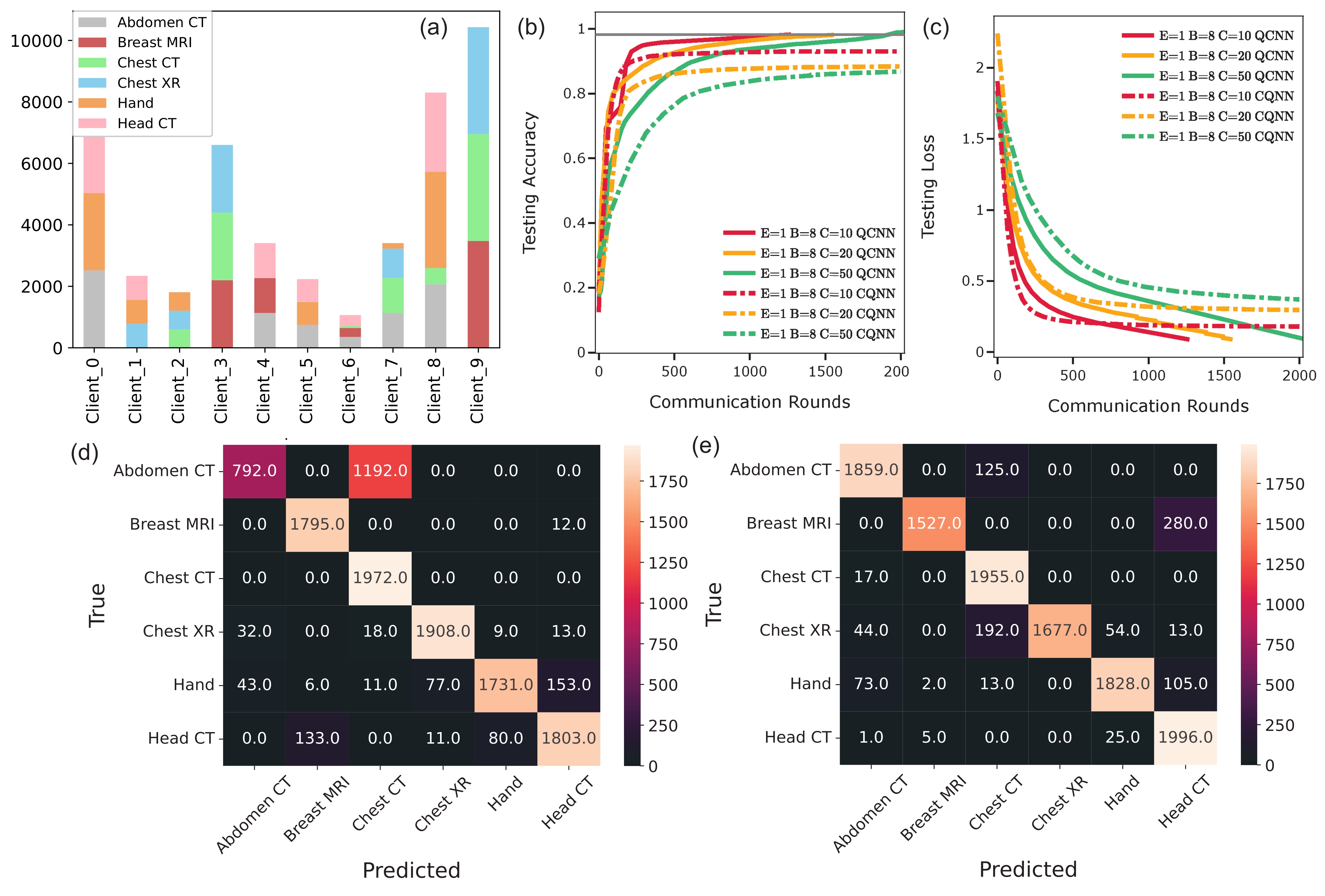}
\caption{A performance comparison between federated quanvolutional QCNN and classical-quantum CQNN on the non-IID2 partition of the MedNIST dataset is illustrated. (a) The non-IID2 distribution shows that each of the 10 clients receives random training samples from at least three different classes. (b) The plot depicts testing accuracy versus communication rounds for the Non-IID2 MedNIST dataset with 10, 20, and 50 clients. (c) The validation loss curves indicate that the CQNN model converges faster and more smoothly. (d-e)  Confusion matrix for the testing set after training for 100 rounds, and finally classified the relevant classes \cite{139}.}
 \label{niid_results}
\end{figure*}

In recent years, the concept of Quantum Federated Learning (QFL) has been proposed, with applications in areas such as healthcare and drug discovery, where QML models are applied in a distributed manner. QFL enables collaborative learning across medical data distributed among multiple hospitals, while preserving data privacy by avoiding direct data sharing. By combining the computational advantages of quantum models with the federated learning framework, QFL allows institutions to train more accurate and robust models, leverage diverse datasets, and reduce the risk of data leakage. Furthermore, QFL has the potential to accelerate training on high-dimensional data, improve generalization across heterogeneous sources, and provide a scalable paradigm for privacy-preserving collaborative AI in sensitive domains. 

Fig.~\ref{niid_results} analyzes the behavior of federated quanvolutional QCNN and hybrid classical–quantum CNN (CQNN) models under a challenging non-IID2 data partition of the MedNIST dataset. In this setting, the training data are heterogeneously distributed such that each of the participating clients receives samples drawn from multiple, randomly selected classes, reflecting realistic data imbalance across decentralized healthcare institutions. This partitioning strategy introduces statistical heterogeneity while maintaining partial class overlap among clients, allowing for a more rigorous evaluation of federated model robustness.

The evolution of testing accuracy as a function of communication rounds demonstrates the impact of client count on convergence behavior for both models. Experiments with 10, 20, and 50 clients reveal that increasing the number of participating clients generally slows convergence due to greater data fragmentation, while the overall accuracy trend remains stable. In parallel, the validation loss trajectories highlight differences in the optimization dynamics, with the CQNN model converging more smoothly and faster than the federated QCNN. This suggests that classical–quantum hybrid architectures may benefit from improved gradient stability under non-IID conditions.

Finally, the confusion matrices obtained after 100 communication rounds provide a class-wise assessment of classification performance on the testing set. These results confirm that both models correctly identify the relevant MedNIST classes despite data heterogeneity, while also revealing residual misclassification patterns associated with visually similar medical imaging categories. Overall, the figure demonstrates the feasibility of federated quantum and hybrid learning approaches under non-IID data distributions and highlights trade-offs between convergence speed and robustness in distributed medical image classification tasks \cite{139}.

Several studies have explored quantum-enhanced models such as quanvolutional neural networks, quantum convolutional neural networks, and variational quantum circuits to improve data representation and learning efficiency in privacy-sensitive medical environments. In parallel, federated quantum machine learning approaches have been proposed to enable collaborative model training across institutions without sharing raw clinical data, while quantum tensor networks have demonstrated efficient and compact representations for large-scale medical datasets. Collectively, these developments underscore the potential of quantum technologies to address key challenges in healthcare, including data privacy preservation, scalability across institutions, and predictive accuracy. These scenarios illustrate how QFL can securely process heterogeneous and sensitive healthcare data while maintaining patient confidentiality.

 In 2024, researchers employed variational quantum circuits (VQCs) to develop a QFL framework capable of predicting coronary heart disease, diagnosing diabetes, and distinguishing between malignant and benign cancer under non-IID data distributions across institutions. This work also showed that applying quantum natural gradient descent (QNGD) optimization could significantly reduce communication rounds while enhancing model performance \cite{135}. In the same year, quantum tensor networks (QTNs) were introduced and validated on large-scale medical imaging datasets, including chest radiographs (RSNA and NIH), MRI brain scans (ADNI), and CT-kidney datasets. Results indicated that QTNs with higher qubit entanglement achieved superior classification accuracy compared to many state-of-the-art federated learning methods, while requiring fewer parameters \cite{136}.

Earlier, in 2023, quantum convolutional neural networks (QCNNs) were trained in a distributed manner across edge devices, with evaluations performed on PneumoniaMNIST and CT kidney disease datasets. The study achieved accuracies of 86.3\% for pneumonia and 92.8\% for CT kidney disease, comparable to classical CNNs, while also introducing an efficient client selection mechanism to reduce communication overhead \cite{137,138}. Similarly, a hybrid quanvolutional neural network (QNN) was proposed to enable distributed learning without direct data exchange. Tested on COVID-19 and MedNIST datasets, this method demonstrated a reduction in communication rounds by 2–39\% compared to federated stochastic gradient descent, while maintaining competitive predictive performance \cite{139}.

QFL provides a powerful framework for addressing key challenges in healthcare analytics by combining distributed machine learning with near-term quantum computing. The approach enables privacy-preserving training of quantum–classical models directly on sensitive medical data, allowing accurate diagnosis, detection, and predictive modeling from medical imaging and clinical datasets without requiring raw data to leave institutional boundaries. By coupling quantum simulations with machine-learning workflows, QFL can capture complex, high-dimensional correlations that classical models struggle to represent, thereby enhancing predictive accuracy and robustness.

QFL distributes computational workloads across multiple quantum and hybrid quantum–classical devices, improving training efficiency while reducing centralized computational bottlenecks. This distributed paradigm supports collaborative learning across hospitals and research centers, allowing institutions to benefit from shared model improvements without exposing patient-level information. As a result, real-time and quantitative predictive insights can be generated while maintaining strict data governance and compliance with healthcare privacy regulations.

  \begin{figure*}[!ht]
	\centering
\includegraphics[scale=0.75]{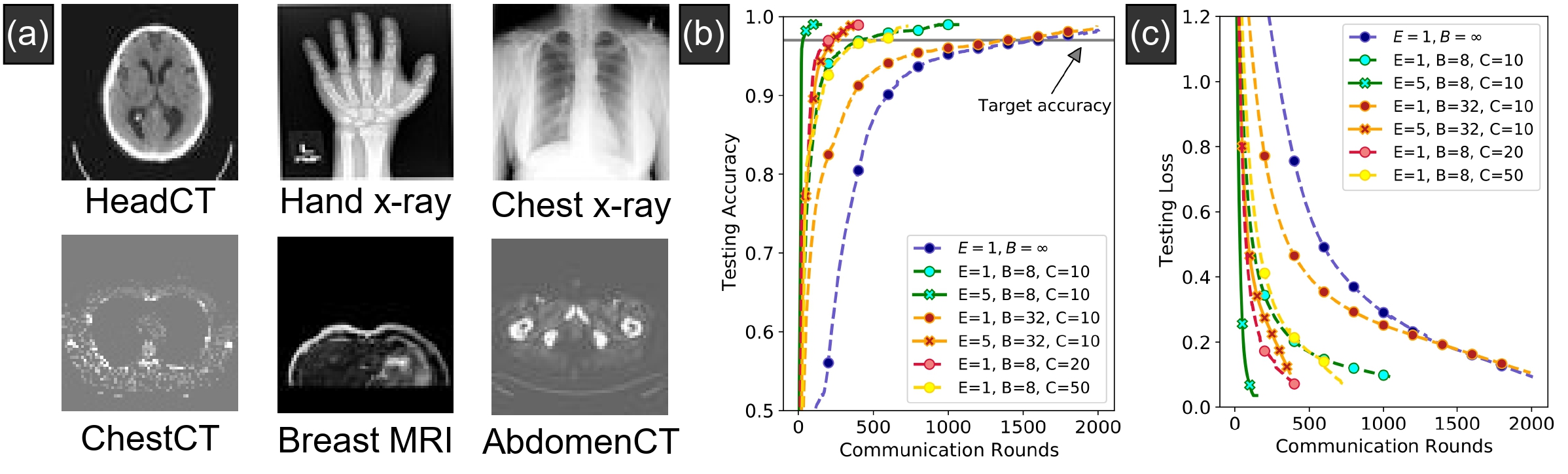}
\caption{(a) Representative medical images from the MedNIST dataset covering six classes: HeadCT, Hand X-ray, Chest X-ray, ChestCT, BreastMRI, and AbdomenCT. (b–c) Classification performance of the global model on a non-IID partition of MedNIST, illustrating the behavior of federated QCNN under varying local epochs and batch sizes. Experiments are conducted with 10, 20, and 50 clients. The gray horizontal line denotes the target accuracy of 97\%. Increasing the number of local epochs (\textit{E}) or decreasing the batch size (\textit{B}) enables the federated quantum model to achieve the target accuracy in fewer communication rounds \cite{139}. Under non-IID data distributions, a speedup of up to 39.7$\times$ is observed through appropriate optimization of batch size and local training epochs.}
 \label{mednist_results}
\end{figure*}

Fig~\ref{mednist_results} illustrates the experimental evaluation of the federated quantum convolutional neural network (QCNN) on the MedNIST dataset under non-IID data distributions. Panel (a) presents representative medical images from the six MedNIST classes, highlighting the diversity of imaging modalities considered in this study. Panels (b) and (c) report the classification performance of the federated global model when the dataset is partitioned across 10, 20, and 50 clients, demonstrating the impact of varying local epochs and batch sizes on convergence behavior. The horizontal gray line indicates the target accuracy of 97\%, corresponding to centralized training without data partitioning. The results show that increasing the number of local training epochs or reducing the batch size enables the federated quantum model to reach the target accuracy in fewer communication rounds. Notably, under non-IID settings, appropriately optimizing these parameters leads to a substantial acceleration of convergence, achieving a speedup of up to 39.7$\times$, thereby highlighting the efficiency and scalability of the federated QCNN framework in distributed medical imaging scenarios \cite{139}.

The framework further incorporates privacy-preserving encryption techniques and post-quantum cryptographic safeguards, providing resilience against both classical and quantum adversaries. By integrating secure aggregation, encrypted parameter exchange, and quantum-resistant security mechanisms, QFL strengthens trust in distributed healthcare AI systems operating in adversarial environments. The overall system is designed to be rigorously benchmarked against state-of-the-art classical federated and centralized learning approaches, demonstrating tangible gains in accuracy, computational efficiency, scalability, and security resilience.

\begin{figure*}[!ht]
	\centering
\includegraphics[scale=0.55]{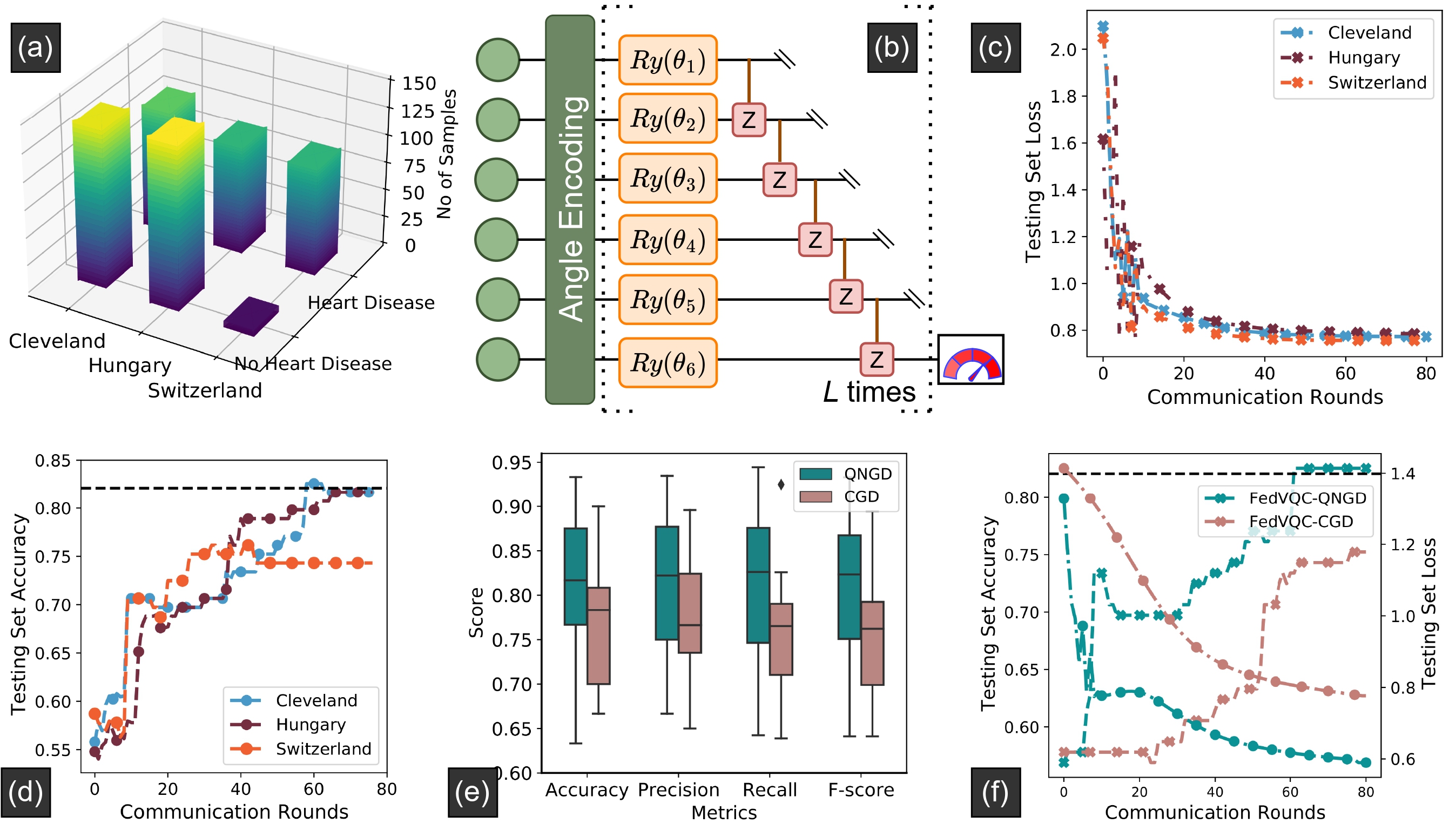}
\caption{(a) A three-dimensional bar chart illustrating binary classification outcomes for Heart Disease and No Heart Disease categories, together with the corresponding sample contributions from the Cleveland, Hungary, and Switzerland hospital datasets. (b) A 13-qubit variational quantum circuit employing angle encoding via Ry rotations and a linear entanglement pattern. (c–d) Testing loss and testing accuracy curves demonstrating the performance of locally trained models from the three hospitals on the Heart Disease (HD) dataset. The black-dotted horizontal line denotes the target accuracy achieved by training on the full dataset without federated learning. (e–f) Evaluation of the performance and robustness of the FedVQC model across different datasets using quantum natural gradient descent (QNGD) and classical gradient descent (CGD) optimizers. Model performance is assessed using standard classification metrics, including accuracy, precision, recall, and F1-score, on the test set, with visual comparisons of testing accuracy and loss for both QNGD- and CGD-optimized FedVQC models \cite{135}.}
 \label{heart_results}
\end{figure*}

Fig.~\ref{heart_results} presents a comprehensive analysis of the proposed federated variational quantum circuit (FedVQC) framework applied to heart disease classification. Panel (a) visualizes the binary classification outcomes for Heart Disease and No Heart Disease using a three-dimensional bar plot, together with the distribution of samples contributed by the Cleveland, Hungary, and Switzerland hospital datasets. This representation emphasizes the heterogeneity and imbalance inherent in real-world clinical data across institutions. Panel (b) depicts the structure of the 13-qubit variational quantum circuit employed in the study, which utilizes angle encoding through Ry rotations combined with a linear entanglement topology. Panels (c) and (d) show the testing loss and testing accuracy curves for the locally trained models at each hospital, providing insight into individual client performance during federated training. The black dotted horizontal line denotes the target accuracy obtained from centralized training on the full dataset, serving as a benchmark for evaluating federated learning performance. Panels (e) and (f) compare the robustness and predictive capability of the FedVQC model across different datasets using quantum natural gradient descent (QNGD) and classical gradient descent (CGD) optimizers. Model performance is assessed using standard classification metrics, including accuracy, precision, recall, and F1-score, along with visualizations of testing accuracy and loss, demonstrating the advantages of QNGD-based optimization in federated quantum learning settings \cite{135}.

Recently, a quantum federated learning (QFL) framework was proposed that uses Fisher information-theoretic measures derived from local quantum models to guide parameter aggregation in the presence of heterogeneous data\cite{140}. By emphasizing parameters that most strongly affect model performance, this approach enhances robustness and improves training efficiency. Experiments on the ADNI and MNIST datasets demonstrate that our method outperforms quantum federated averaging in terms of accuracy and stability\cite{140}.

By harnessing the computational advantages of quantum computing within a federated learning paradigm, QFL enables collaborative model training while preserving data privacy across distributed healthcare and pharmaceutical institutions. The decentralized nature of QFL facilitates robust and generalizable learning by capturing heterogeneity across patient populations, imaging modalities, and clinical environments. Beyond medical imaging, QFL has the potential to transform a wide range of pharmaceutical applications, including drug discovery and development, personalized medicine, clinical trial optimization, supply chain management, pharmacovigilance, and collaborative drug pricing strategies. Together, these capabilities position QFL as a promising framework for secure, scalable, and next-generation data-driven healthcare analytics. This paradigm offers a viable pathway toward integrating quantum advantages with real-world healthcare workflows under stringent privacy and regulatory constraints.

\section{Conclusion}
\label{sec:conclusion}

This Review illustrates the ongoing research in quantum machine learning that is particularly relevant to complex systems, emphasizing approaches that address practical challenges in representation, sampling, learning dynamics, and deployment.
We provide an account of how QML is evolving from a collection of conceptual proposals into a framework capable of engaging with realistic scientific and data-driven problems.
We first examined quantum-enabled variational Monte Carlo strategies for training neural-network quantum states, highlighting how quantum resources can be integrated into classical learning loops to mitigate sampling bottlenecks in strongly correlated systems.
We then reviewed emerging efforts to understand the learning dynamics of QML models, where tools from quantum information and many-body physics provide insight into optimization landscapes, trainability, and stability beyond conventional performance metrics.
Together, these developments underscore that scalable QML requires not only expressive models but also principled mechanisms for efficient training and diagnostic understanding.
Building on this methodological foundation, we surveyed application-driven studies in domains such as drug discovery, cancer biology, and agro-climate modeling, where high-dimensional structure, limited data, and domain constraints pose stringent challenges for learning algorithms.
These applications serve both as motivation and benchmarks for QML, revealing the conditions under which hybrid quantum-classical approaches may offer practical advantages. Finally, we discuss federated quantum machine learning as a pathway toward distributed and privacy preserving learning, particularly relevant for domains such as healthcare, materials science, and large-scale scientific collaborations where data sharing is restricted. In the context of materials design, federated QML enables multiple laboratories or industrial partners to collaboratively train quantum models on locally generated simulation or experimental data, such as electronic structure calculations, spectroscopy, or synthesis outcomes, without exposing proprietary datasets. By aggregating learned quantum features rather than raw data, this approach can accelerate the discovery of functional materials, optimize design pipelines, and support secure multi-institutional innovation while respecting intellectual property and data governance constraints.

Looking ahead, the continued development of QML will depend on progress on multiple fronts, including improved quantum hardware, quantum error correction, scalable training protocols, and system-level integration with classical infrastructure. Equally important will be a deeper theoretical understanding of when and why quantum resources improve learning performance for complex systems.
By consolidating insights across algorithms, theory, and applications, this Review aims to clarify both the opportunities and the limitations of quantum machine learning, and to help guide future efforts toward robust and impactful quantum-enabled learning systems.



\section*{Acknowledgments}
The authors would like to acknowledge funding from the U.S. Department of Energy (DOE) Office of Basic Energy Sciences, Award No. DE-SC0026309.

\bibliographystyle{plain}
\bibliography{ref}

@article{medvidovic2024neural,
  title={Neural-network quantum states for many-body physics},
  author={Medvidovi{\'c}, Matija and Moreno, Javier Robledo},
  journal={arXiv preprint arXiv:2402.11014},
  year={2024}
}

@article{carleo2017solving,
  title={Solving the quantum many-body problem with artificial neural networks},
  author={Carleo, Giuseppe and Troyer, Matthias},
  journal={Science},
  volume={355},
  number={6325},
  pages={602--606},
  year={2017},
  publisher={American Association for the Advancement of Science}
}

@article{McClean2018BarrenPlateau,
  author  = {McClean, Jarrod R. and others},
  title   = {Barren Plateaus in Quantum Neural Network Training Landscapes},
  journal = {Nature Communications},
  year    = {2018},
  volume  = {9},
  pages   = {4812},
  doi     = {10.1038/s41467-018-07090-4}
}

@article{Peruzzo2014VQE,
  author  = {Peruzzo, Alberto and others},
  title   = {A Variational Eigenvalue Solver on a Photonic Quantum Processor},
  journal = {Nature Communications},
  year    = {2014},
  volume  = {5},
  pages   = {4213},
  doi     = {10.1038/ncomms5213}
}

@article{biamonte2017quantum,
  title={Quantum machine learning},
  author={Biamonte, Jacob and Wittek, Peter and Pancotti, Nicola and Rebentrost, Patrick and Wiebe, Nathan and Lloyd, Seth},
  journal={Nature},
  volume={549},
  number={7671},
  pages={195--202},
  year={2017},
  publisher={Nature Publishing Group UK London}
}

@article{cerezo2021variational,
  title={Variational quantum algorithms},
  author={Cerezo, Marco and Arrasmith, Andrew and Babbush, Ryan and Benjamin, Simon C and Endo, Suguru and Fujii, Keisuke and McClean, Jarrod R and Mitarai, Kosuke and Yuan, Xiao and Cincio, Lukasz and others},
  journal={Nature Reviews Physics},
  volume={3},
  number={9},
  pages={625--644},
  year={2021},
  publisher={Nature Publishing Group UK London}
}

@article{sureshbabu2021implementation,
  title={Implementation of Quantum Machine Learning for Electronic Structure Calculations of Periodic Systems on Quantum Computing Devices},
  author={Sureshbabu, Shree Hari and Sajjan, Manas and Oh, Sangchul and Kais, Sabre},
  journal={Journal of Chemical Information and Modeling},
  year={2021},
  publisher={ACS Publications}
}

@article{sajjan2022quantum,
  title={Quantum machine learning for chemistry and physics},
  author={Sajjan, Manas and Li, Junxu and Selvarajan, Raja and Sureshbabu, Shree Hari and Kale, Sumit Suresh and Gupta, Rishabh and Singh, Vinit and Kais, Sabre},
  journal={Chemical Society Reviews},
  year={2022},
  publisher={Royal Society of Chemistry}
}

@article{havlivcek2019supervised,
  title={Supervised learning with quantum-enhanced feature spaces},
  author={Havl{\'\i}{\v{c}}ek, Vojt{\v{e}}ch and C{\'o}rcoles, Antonio D and Temme, Kristan and Harrow, Aram W and Kandala, Abhinav and Chow, Jerry M and Gambetta, Jay M},
  journal={Nature},
  volume={567},
  number={7747},
  pages={209--212},
  year={2019},
  publisher={Nature Publishing Group}
}

@article{sajjan2023imaginary,
  title={Imaginary components of out-of-time-order correlator and information scrambling for navigating the learning landscape of a quantum machine learning model},
  author={Sajjan, Manas and Singh, Vinit and Selvarajan, Raja and Kais, Sabre},
  journal={Physical Review Research},
  volume={5},
  number={1},
  pages={013146},
  year={2023},
  publisher={APS}
}

@article{Gutierrez2022Real,
  author       = {Irene L. Gutiérrez and Christian B. Mendl},
  title        = {Real time evolution with neural‐network quantum states},
  journal      = {Quantum},
  volume       = {6},
  pages        = {627},
  year         = {2022},
  doi          = {10.22331/Q-2022-01-20-627}
}

@article{Carleo2016Solving,
  author = {Carleo, Giuseppe and Troyer, Matthias},
  title = {Solving the quantum many-body problem with artificial neural networks},
  journal = {arXiv preprint arXiv:1606.02318},
  year = {2016}
}

@article{Yang2019Approximating,
  author = {Yang, Yichen},
  title = {Approximating Ground States by Neural Network Quantum States},
  journal = {arXiv preprint arXiv:1905.00000},
  year = {2019}
}

@article{Kim2024Neural,
  author = {Kim, J. and others},
  title = {Neural-network quantum states for ultra-cold Fermi gases},
  journal = {Communications Physics},
  year = {2024},
  volume = {7},
  pages = {41},
  doi = {10.1038/s42005-024-01613-w}
}

@article{Schuld2019Neural,
  author = {Schuld, M. and others},
  title = {Neural Networks Take on Open Quantum Systems},
  journal = {Physics},
  year = {2019},
  volume = {12},
  pages = {74},
  doi = {10.1103/Physics.12.74}
}

@book{Schuld2018Supervised,
  author    = {Schuld, Maria and Petruccione, Francesco},
  title     = {Supervised Learning with Quantum Computers},
  publisher = {Springer},
  year      = {2018}
}

@article{Ciliberto2018QMLPerspective,
  author  = {Ciliberto, Carlo and others},
  title   = {Quantum Machine Learning: A Classical Perspective},
  journal = {Proceedings of the Royal Society A},
  year    = {2018},
  volume  = {474},
  pages   = {20170551},
  doi     = {10.1098/rspa.2017.0551}
}

@article{Dunjko2018MLReview,
  author  = {Dunjko, Vedran and Briegel, Hans J.},
  title   = {Machine Learning and Artificial Intelligence in the Quantum Domain},
  journal = {Reports on Progress in Physics},
  year    = {2018},
  volume  = {81},
  pages   = {074001},
  doi     = {10.1088/1361-6633/aab406}
}

@article{Melnikov2023Quantum,
  author = {Melnikov, A. and others},
  title = {Quantum machine learning: from physics to software},
  journal = {Quantum Studies: Mathematics and Foundations},
  year = {2023},
  volume = {10},
  number = {4},
  pages = {343–357},
  doi = {10.1080/23746149.2023.2165452}
}

@article{Matthew2025Quantum,
  author = {Bamidele Matthew and Karkala, Sudarshana and Hossain, Sazzad and Krishnapatnam, Mahendra and Aggarwal, Ankur and Zahir, Zarif and Pandhare, Harshad Vijay and Shah, Varun},
  title = {Quantum Machine Learning for Material Science Simulations: Opportunities and Challenges},
  journal = {arXiv preprint arXiv:2506.12345},
  year = {2025}
}

@article{chemicalAI2024Explainable,
  author = {Gallegos, M. and others},
  title = {Explainable chemical artificial intelligence from accurate neural networks},
  journal = {Journal of Chemical Physics},
  year = {2024},
  volume = {161},
  pages = {204123},
  doi = {10.1063/5.0150000}
}

@article{Sajjan2023Polynomial,
  author    = {Sajjan, Manas and Singh, Vinit and Kais, Sabre},
  title     = {Polynomially Efficient Quantum-Enabled Variational Monte Carlo for Training Neural-Network Quantum States for Physico-Chemical Applications},
  journal   = {arXiv preprint arXiv:2307.XXXXX},
  year      = {2023},
  note      = {In preparation / submitted},
}

@article{Sajjan2023ImaginaryOTOC,
  author    = {Sajjan, Manas and Selvarajan, Raja and Oh, Sangchul and Kais, Sabre},
  title     = {Imaginary Components of Out-of-Time-Order Correlators and Information Scrambling for Navigating the Learning Landscape of a Quantum Machine Learning Model},
  journal   = {Physical Review Research},
  year      = {2023},
  volume    = {5},
  number    = {1},
  pages     = {013146},
  doi       = {10.1103/PhysRevResearch.5.013146}
}

@article{wolf2008area,
  title={Area laws in quantum systems: mutual information and correlations},
  author={Wolf, Michael M and Verstraete, Frank and Hastings, Matthew B and Cirac, J Ignacio},
  journal={Physical review letters},
  volume={100},
  number={7},
  pages={070502},
  year={2008},
  publisher={APS}
}

@article{bhatia2023quantum,
  title={Quantum machine learning predicting ADME-Tox properties in drug discovery},
  author={Bhatia, Amandeep Singh and Saggi, Mandeep Kaur and Kais, Sabre},
  journal={Journal of Chemical Information and Modeling},
  volume={63},
  number={21},
  pages={6476--6486},
  year={2023},
  publisher={ACS Publications}
}

@article{kais2023quantum,
  title   = {Quantum federated learning in healthcare: the shift from development to deployment and from models to data},
  author  = {Kais, Sabre and Bhatia, Amandeep and Alam, Muhammad},
  journal = {Research Square},
  year    = {2023},
  doi     = {10.21203/rs.3.rs-2723753/v1}
}

@article{bhatia2023federated,
  title={Federated quanvolutional neural network: a new paradigm for collaborative quantum learning},
  author={Bhatia, Amandeep Singh and Kais, Sabre and Alam, Muhammad Ashraful},
  journal={Quantum Science and Technology},
  volume={8},
  number={4},
  pages={045032},
  year={2023},
  publisher={IOP Publishing}
}

@article{bhatia2023handling,
  title={Handling privacy-sensitive clinical data with federated quantum machine learning},
  author={Bhatia, Amandeep and Kais, Sabre and Alam, Muhammad},
  journal={Bulletin Of The American Physical Society},
  volume={68},
  year={2023},
  publisher={APS}
}

@article{saggi2026multi,
  title={Multi-Omic and quantum machine learning integration for lung subtypes classification},
  author={Saggi, Mandeep Kaur and Bhatia, Amandeep Singh and Mensah, Isaiah K and Gowher, Humaira and Kais, Sabre},
  journal={Future Generation Computer Systems},
  volume={174},
  pages={107905},
  year={2026},
  publisher={Elsevier}
}

@inproceedings{saggi2024mqml,
  title={MQML: Multi-Omic Quantum Machine Learning Based Cancer Classification, Biomarker Identification in Human Lung Adenocarcinoma},
  author={Saggi, Mandeep Kaur and Kais, Sabre},
  booktitle={2024 IEEE International Conference on Quantum Computing and Engineering (QCE)},
  volume={1},
  pages={1713--1720},
  year={2024},
  organization={IEEE}
}

@article{isik2025multimodal,
  title={Multimodal Quantum Vision Transformer for Enzyme Commission Classification from Biochemical Representations},
  author={Isik, Murat and Saggi, Mandeep Kaur and Gowher, Humaira and Kais, Sabre},
  journal={arXiv preprint arXiv:2508.14844},
  year={2025}
}

@article{saggi2023using,
  title={Using quantum circuits with convolutional neural networks for multi-object detection and classification},
  author={Saggi, Mandeep and Kais, Sabre},
  journal={Bulletin of the American Physical Society},
  volume={68},
  year={2023},
  publisher={APS}
}

@inproceedings{bhatia2024robustness,
  title={Robustness of Quantum Federated Learning (QFL) Against “Label Flipping Attacks” for Lithography Hotspot Detection in Semiconductor Manufacturing},
  author={Bhatia, Amandeep Singh and Kais, Sabre and Alam, Muhammad Ashraful},
  booktitle={2024 IEEE international reliability physics symposium (IRPS)},
  pages={1--4},
  year={2024},
  organization={IEEE}
}

@article{brandhofer2022benchmarking,
  title={Benchmarking the performance of portfolio optimization with QAOA: S. Brandhofer et al.},
  author={Brandhofer, Sebastian and Braun, Daniel and Dehn, Vanessa and Hellstern, Gerhard and H{\"u}ls, Matthias and Ji, Yanjun and Polian, Ilia and Bhatia, Amandeep Singh and Wellens, Thomas},
  journal={Quantum Information Processing},
  volume={22},
  number={1},
  pages={25},
  year={2022},
  publisher={Springer}
}

@article{selick2002emerging,
  title={The emerging importance of predictive ADME simulation in drug discovery},
  author={Selick, Harold E and Beresford, Alan P and Tarbit, Michael H},
  journal={Drug Discovery Today},
  volume={7},
  number={2},
  pages={109--116},
  year={2002},
  publisher={Elsevier}
}

@article{maltarollo2015applying,
  title={Applying machine learning techniques for ADME-Tox prediction: a review},
  author={Maltarollo, Vin{\'\i}cius Gon{\c{c}}alves and Gertrudes, Jadson Castro and Oliveira, Patr{\'\i}cia Rufino and Honorio, Kathia Maria},
  journal={Expert opinion on drug metabolism \& toxicology},
  volume={11},
  number={2},
  pages={259--271},
  year={2015},
  publisher={Taylor \& Francis}
}

@article{batra2021quantum,
  title={Quantum machine learning algorithms for drug discovery applications},
  author={Batra, Kushal and Zorn, Kimberley M and Foil, Daniel H and Minerali, Eni and Gawriljuk, Victor O and Lane, Thomas R and Ekins, Sean},
  journal={Journal of chemical information and modeling},
  volume={61},
  number={6},
  pages={2641--2647},
  year={2021},
  publisher={ACS Publications}
}

@article{huang2021therapeutics,
  title={Therapeutics data commons: Machine learning datasets and tasks for drug discovery and development},
  author={Huang, Kexin and Fu, Tianfan and Gao, Wenhao and Zhao, Yue and Roohani, Yusuf and Leskovec, Jure and Coley, Connor W and Xiao, Cao and Sun, Jimeng and Zitnik, Marinka},
  journal={arXiv preprint arXiv:2102.09548},
  year={2021}
}

@ARTICLE{2025arXiv250517756S,
       author = {{Sahin}, M. Emre and {Altamura}, Edoardo and {Wallis}, Oscar and {Wood}, Stephen P. and {Dekusar}, Anton and {Millar}, Declan A. and {Imamichi}, Takashi and {Matsuo}, Atsushi and {Mensa}, Stefano},
        title = "{Qiskit Machine Learning: an open-source library for quantum machine learning tasks at scale on quantum hardware and classical simulators}",
      journal = {arXiv e-prints},
         year = 2025,
        month = may,
          eid = {arXiv:2505.17756},
        pages = {arXiv:2505.17756},
archivePrefix = {arXiv},
       eprint = {2505.17756},
 primaryClass = {quant-ph},
       adsurl = {https://ui.adsabs.harvard.edu/abs/2025arXiv250517756S}
}

@article{robertson2023squeeze,
  title={sQueeze: Accelerated Quantum Pulse Schedules},
  author={Robertson, Lilian Hunt Alan},
  journal={arXiv preprint arXiv:2311.08742},
  year={2023}
}

@article{yadav2025knowledge,
  title={Knowledge mapping of quantum computing for climate change research and sustainable innovation},
  author={Yadav, Mamta and Rawat, Keshav Singh},
  journal={Quality \& Quantity},
  pages={1--34},
  year={2025},
  publisher={Springer}
}

@article{bhatia2018quantifying,
  title={Quantifying matrix product state},
  author={Bhatia, Amandeep Singh and Kumar, Ajay},
  journal={Quantum Information Processing},
  volume={17},
  number={3},
  pages={41},
  year={2018},
  publisher={Springer}
}

@article{bhatia2019matrix,
  title={Matrix product state--based quantum classifier},
  author={Bhatia, Amandeep Singh and Saggi, Mandeep Kaur and Kumar, Ajay and Jain, Sushma},
  journal={Neural computation},
  volume={31},
  number={7},
  pages={1499--1517},
  year={2019},
  publisher={MIT Press One Rogers Street, Cambridge, MA 02142-1209, USA journals-info~…}
}

@article{saggi2019reference,
  title={Reference evapotranspiration estimation and modeling of the Punjab Northern India using deep learning},
  author={Saggi, Mandeep Kaur and Jain, Sushma},
  journal={Computers and Electronics in Agriculture},
  volume={156},
  pages={387--398},
  year={2019},
  publisher={Elsevier}
}

@article{saggi2022survey,
  title={A Survey Towards Decision Support System on Smart Irrigation Scheduling Using Machine Learning approaches: KS Saggi, S. Jain},
  author={Saggi, Mandeep Kaur and Jain, Sushma},
  journal={Archives of computational methods in engineering},
  volume={29},
  number={6},
  pages={4455--4478},
  year={2022},
  publisher={Springer}
}

@article{huggins2019towards,
  title={Towards quantum machine learning with tensor networks},
  author={Huggins, William and Patil, Piyush and Mitchell, Bradley and Whaley, K Birgitta and Stoudenmire, E Miles},
  journal={Quantum Science and technology},
  volume={4},
  number={2},
  pages={024001},
  year={2019},
  publisher={IOP Publishing}
}

@article{stoudenmire2016supervised,
  title={Supervised learning with tensor networks},
  author={Stoudenmire, Edwin and Schwab, David J},
  journal={Advances in neural information processing systems},
  volume={29},
  year={2016}
}

@inproceedings{135,
  title={Communication-efficient Quantum Federated Learning Optimization for Multi-Center Healthcare Data},
  author={Bhatia, A. and Saggi, M. and Kais, S.},
  booktitle={IEEE-EMBS International Conference on Biomedical and Health Informatics},
  year={2024},
  publisher={IEEE}
}

@article{136,
  title={Federated learning with tensor networks: a quantum AI framework for healthcare},
  author={Bhatia, Amandeep Singh and Neira, David E Bernal},
  journal={Machine Learning: Science and Technology},
  volume={5},
  number={4},
  pages={045035},
  year={2024},
  publisher={IOP Publishing}
}

@article{137,
  author  = {Bhatia, Amandeep Singh and Kais, Sabre and Alam, Muhammad Ashraful},
  title   = {Quantum Federated Learning in Healthcare: The Shift from Development to Deployment and from Models to Data},
  journal = {IEEE Journal of Biomedical and Health Informatics},
  year    = {2025},
  month   = aug,
  doi     = {10.1109/JBHI.2025.3596156},
  pmid    = {40768459}
}

@inproceedings{138,
  title={Handling privacy-sensitive clinical data with federated quantum machine learning},
  author={Bhatia, A. and Kais, S. and Alam, M.},
  booktitle={APS March Meeting Abstracts},
  volume={2023},
  pages={T70--007},
  year={2023},
  publisher={American Physical Society}
}

@article{139,
  title={Federated quanvolutional neural network: a new paradigm for collaborative quantum learning},
  author={Bhatia, A. and Kais, S. and Alam, M.},
  journal={Quantum Science and Technology},
  volume={8},
  number={4},
  pages={045032},
  year={2023},
  publisher={IOP Publishing}
}

@inproceedings{140,
  title={Enhancing Quantum Federated Learning with Fisher Information-Based Optimization},
  author={Bhatia, Amandeep Singh and Kais, Sabre},
  booktitle={2025 IEEE International Conference on Quantum Computing and Engineering (QCE)},
  volume={1},
  pages={1015--1020},
  year={2025},
  organization={IEEE}
}

@article{chen2021federated,
  title={Federated quantum machine learning},
  author={Chen, Samuel Yen-Chi and Yoo, Shinjae},
  journal={Entropy},
  volume={23},
  number={4},
  pages={460},
  year={2021},
  publisher={MDPI}
}

@article{Holmes2022Connectivity,
  author  = {Holmes, Zoe and others},
  title   = {Connecting Ansätze Expressibility to Gradient Magnitudes},
  journal = {PRX Quantum},
  year    = {2022},
  volume  = {3},
  pages   = {010313},
  doi     = {10.1103/PRXQuantum.3.010313}
}

@article{Troyer2005ComputationalComplexity,
  author  = {Troyer, Matthias and Wiese, Uwe-Jens},
  title   = {Computational Complexity and Fundamental Limitations to Fermionic Quantum Monte Carlo Simulations},
  journal = {Physical Review Letters},
  year    = {2005},
  volume  = {94},
  pages   = {170201},
  doi     = {10.1103/PhysRevLett.94.170201}
}

@article{Binder1986MonteCarlo,
  author  = {Binder, Kurt and Heermann, Dieter},
  title   = {Monte Carlo Simulation in Statistical Physics},
  journal = {Springer},
  year    = {1986}
}

@article{Rudin2019MLRisks,
  author  = {Rudin, Cynthia},
  title   = {Stop Explaining Black Box Machine Learning Models for High Stakes Decisions},
  journal = {Nature Machine Intelligence},
  year    = {2019},
  volume  = {1},
  pages   = {206--215},
  doi     = {10.1038/s42256-019-0048-x}
}

@article{Reichstein2019DeepLearningClimate,
  author  = {Reichstein, Markus and others},
  title   = {Deep Learning and Process Understanding for Data-Driven Earth System Science},
  journal = {Nature},
  year    = {2019},
  volume  = {566},
  pages   = {195--204},
  doi     = {10.1038/s41586-019-0912-1}
}

@article{Esteva2019HealthcareAI,
  author  = {Esteva, Andre and others},
  title   = {A Guide to Deep Learning in Healthcare},
  journal = {Nature Medicine},
  year    = {2019},
  volume  = {25},
  pages   = {24--29},
  doi     = {10.1038/s41591-018-0316-z}
}

@article{Donatella2023AutoregressiveDynamics,
  author    = {Donatella, K. and Vicentini, F. and Carleo, G.},
  title     = {Dynamics with Autoregressive Neural Quantum States},
  journal   = {Physical Review A},
  year      = {2023},
  volume    = {108},
  pages     = {022210},
  doi       = {10.1103/PhysRevA.108.022210}
}

@article{Yuan2019Theory,
  author    = {Yuan, Xiao and Endo, Suguru and Zhao, Qi and Li, Ying and Benjamin, Simon C.},
  title     = {Theory of Variational Quantum Simulation},
  journal   = {Quantum},
  year      = {2019},
  volume    = {3},
  pages     = {191},
  doi       = {10.22331/q-2019-10-07-191}
}

@article{Endo2020GeneralProcesses,
  author    = {Endo, Suguru and Sun, Jinzhao and Li, Ying and Benjamin, Simon C. and Yuan, Xiao},
  title     = {Variational Quantum Simulation of General Processes},
  journal   = {Physical Review Letters},
  year      = {2020},
  volume    = {125},
  number    = {1},
  pages     = {010501},
  doi       = {10.1103/PhysRevLett.125.010501}
}

@article{Vicentini2019OpenSteadyStates,
  author    = {Vicentini, Filippo and Biella, Alberto and Regnault, Nicolas and Carleo, Giuseppe},
  title     = {Variational Neural-Network Ansatz for Steady States in Open Quantum Systems},
  journal   = {Physical Review Letters},
  year      = {2019},
  volume    = {122},
  number    = {25},
  pages     = {250503},
  doi       = {10.1103/PhysRevLett.122.250503}
}

@article{Nagy2019OpenVMC,
  author    = {Nagy, Attila and Savona, Vincenzo},
  title     = {Variational Quantum Monte Carlo Method with a Neural-Network Ansatz for Open Quantum Systems},
  journal   = {Physical Review Letters},
  year      = {2019},
  volume    = {122},
  number    = {25},
  pages     = {250501},
  doi       = {10.1103/PhysRevLett.122.250501}
}

@article{Liu2021OpenSteadyStates,
  author    = {Liu, Jin-Guo and Zhang, Lei and Wang, Lei},
  title     = {Variational Quantum Algorithms for the Steady States of Open Quantum Systems},
  journal   = {Chinese Physics Letters},
  year      = {2021},
  volume    = {38},
  number    = {8},
  pages     = {080301},
  doi       = {10.1088/0256-307X/38/8/080301}
}

@article{Chen2024AdaptiveOpen,
  author    = {Chen, Xiang and Yuan, Xiao and Benjamin, Simon C.},
  title     = {Adaptive Variational Quantum Simulation for Open Quantum Systems},
  journal   = {Quantum},
  year      = {2024},
  volume    = {8},
  pages     = {1252},
  doi       = {10.22331/q-2024-02-13-1252}
}

@article{Watad2023LindbladVQA,
  author    = {Watad, Ahmad and Cohen, Guy and Shacham, Tomer},
  title     = {Variational Quantum Algorithms for Simulation of Lindblad Master Equations},
  journal   = {arXiv preprint arXiv:2305.02815},
  year      = {2023},
  eprint    = {2305.02815},
  archivePrefix = {arXiv},
  primaryClass  = {quant-ph}
}

@article{Higgott2019VQD,
  author    = {Higgott, Oscar and Wang, Daochen and Brierley, Stephen},
  title     = {Variational Quantum Computation of Excited States},
  journal   = {Physical Review Letters},
  year      = {2019},
  volume    = {122},
  number    = {14},
  pages     = {140504},
  doi       = {10.1103/PhysRevLett.122.140504}
}

@article{Pfau2020FermiNet,
  author    = {Pfau, David and Spencer, James S. and Matthews, Alexander G. D. G. and Foulkes, W. M. C.},
  title     = {Ab Initio Solution of the Many-Electron Schrödinger Equation with Deep Neural Networks},
  journal   = {Physical Review Research},
  year      = {2020},
  volume    = {2},
  number    = {3},
  pages     = {033429},
  doi       = {10.1103/PhysRevResearch.2.033429}
}

@article{Hermann2020PauliNet,
  author    = {Hermann, Jan and Sch{\"a}tzle, Zeno and No{\'e}, Frank},
  title     = {Deep Neural Network Solution of the Electronic Schrödinger Equation},
  journal   = {Nature Chemistry},
  year      = {2020},
  volume    = {12},
  pages     = {891--897},
  doi       = {10.1038/s41557-020-0544-y}
}

@article{JACS2021,
  author    = {M Sajjan, SH Sureshbabu, S Kais},
  title     = {Quantum machine-learning for eigenstate filtration in two-dimensional materials},
  journal   = {Journal of the American Chemical Society },
  year      = {2021},
  volume    = {143},
  pages     = {18426-18445},
  doi       = {https://doi.org/10.1021/jacs.1c06246}
  }

\end{document}